\documentclass[fleqn,usenatbib]{mnras}
\usepackage{subfig}
\usepackage{newtxtext,newtxmath}

\usepackage[T1]{fontenc}

\DeclareRobustCommand{\VAN}[3]{#2}
\let\VANthebibliography\thebibliography
\def\thebibliography{\DeclareRobustCommand{\VAN}[3]{##3}\VANthebibliography}

\usepackage{graphicx}	
\usepackage{amsmath}	
\usepackage{caption}
\usepackage{array}
\usepackage{makecell}
\usepackage{tabularx}
\usepackage[normalem]{ulem}
\usepackage{bm}

\newcommand{\Msun}{M_\odot}
\newcommand{\Mdot}{\dot{M}}
\newcommand{\Mdotstar}{\dot{M}_\ast}
\newcommand{\Mdotin}{\dot{M}_\mathrm{in}}

\newcommand{\Pdot}{\dot{P}}
\newcommand{\Pdotsecular}{\dot{P}_{\rm secular}}
\newcommand{\Pdotoutburst}{\langle\dot{P}_{\rm outburst}\rangle}

\newcommand{\rin}{r_\mathrm{in}}

\newcommand{\rco}{r_\mathrm{co}}
\newcommand{\rinmax}{r_\mathrm{in,max}}
\newcommand{\rA}{r_\mathrm{A}}
\newcommand{\reta}{r_\eta}
\newcommand{\rxi}{r_\xi}
\newcommand{\Rin}{R_\mathrm{in}}
\newcommand{\RA}{R_\mathrm{A}}
\newcommand{\Reta}{R_\eta}
\newcommand{\Rxi}{R_\xi}
\newcommand{\Rinmax}{R_\mathrm{in,max}}

\newcommand{\Lacc}{L_\mathrm{acc}}

\newcommand{\Lx}{L_\mathrm{X}}

\newcommand{\Gammaacc}{\Gamma_\mathrm{acc}}
\newcommand{\GammaD}{\Gamma_\mathrm{D}}
\newcommand{\Gammadip}{\Gamma_\mathrm{dip}}

\newcommand{\gpers}{g~s$^{-1}$} 
\newcommand{\ergpers}{erg~s$^{-1}$} 
\newcommand{\spers}{s~s$^{-1}$}

\newcommand{\Alfven}{Alfv$\acute{\mathrm{e}}$n~}
\newcommand{\figba}{\hyperref[fig:1751]{\ref*{fig:1751}a}}
\newcommand{\figbb}{\hyperref[fig:1751]{\ref*{fig:1751}b}}
\newcommand{\figbc}{\hyperref[fig:1751]{\ref*{fig:1751}c}}

\newcommand{\figia}{\hyperref[fig:17494]{\ref*{fig:17494}a}}
\newcommand{\figib}{\hyperref[fig:17494]{\ref*{fig:17494}b}}
\newcommand{\figic}{\hyperref[fig:17494]{\ref*{fig:17494}c}}

\newcommand{\figda}{\hyperref[fig:3057]{\ref*{fig:3057}a}}
\newcommand{\figdb}{\hyperref[fig:3057]{\ref*{fig:3057}b}}
\newcommand{\figdc}{\hyperref[fig:3057]{\ref*{fig:3057}c}}

\newcommand{\figbea}{\hyperref[fig:5934]{\ref*{fig:5934}a}}
\newcommand{\figbeb}{\hyperref[fig:5934]{\ref*{fig:5934}b}}
\newcommand{\figbec}{\hyperref[fig:5934]{\ref*{fig:5934}c}}

\title[The torques acting on AMXPs]{The Torques Acting on Accreting Millisecond X-Ray Pulsars in the Outburst and Quiescent States, and During the Long-Term Evolution}

\author[F. Ertuğrul et al.]{
F. Ertuğrul$^{1}$\thanks{E-mail: fatmanur.ertugrul@sabanciuniv.edu},
A. A. Gen\c{c}ali$^{1}$,
\"{U}. Ertan$^{1}$
and N. Niang$^{1}$
\\
$^{1}$Sabanc{\i} University, Orhanl{\i}, Tuzla, 34956, \.{I}stanbul, Turkey
}

\date{Accepted XXX. Received YYY; in original form ZZZ}

\pubyear{\the\year{}}

\begin{document}
\label{firstpage}
\pagerange{\pageref{firstpage}--\pageref{lastpage}}
\maketitle

\begin{abstract}
Accreting millisecond X-ray pulsars (AMXPs) are transient X-ray sources likely to be in the final evolutionary phases of low-mass X-ray binaries (LMXBs). In this work, we have investigated the torque and X-ray luminosity variations of five AMXPs during outburst and quiescence, using a model previously employed to explain some typical behaviours of LMXBs. Most of these systems spin up in the outburst state and spin down in quiescence, while they slow down on the long term. We have obtained reasonable results with the model in agreement with these observations. We have found that the torques produced by the disc-magnetosphere interaction, the magnetic dipole radiation, and the mass accretion are compatible with the X-ray luminosity and rotational properties of the sources in their outburst and quiescent states, and during the resultant long-term evolution. Nevertheless, our results are not sufficient to rule out the spin-down contribution of the gravitational radiation torques due to significant timing noise and uncertainties about the bolometric corrections for X-ray luminosity during the outbursts of some sources.
\end{abstract}




\section{Introduction}\label{intro}

Low-mass X-ray binaries (LMXBs) are binary star systems that contain a neutron star (NS) or a black hole, accreting matter from a low-mass companion star ($<\Msun$) via Roche-lobe overflow \citep{Frank2002}. Hereafter, the term LMXB refers to the systems harbouring an NS. LMXBs become transient sources when the average mass accretion rate decreases below a critical value, which depends on the orbital period \citep{meyer1984,van1994,Dubus1999}. Transient LMXBs exhibit periodic outbursts lasting days to months, separated by long quiescent intervals varying from months to years, during which little or no accretion occurs \citep{Bahramian2022}. These outbursts are triggered by thermal-viscous instabilities \citep[see e.g.][for details]{Dubus1999,Lasota2001,Frank2002}.

Accreting millisecond X-ray pulsars (AMXPs), which exhibit coherent X-ray pulses with spin periods, $P$, in the millisecond range, are a subclass of LMXBs. All known AMXPs are transient sources, and their X-ray pulsations have been observed only during outburst states \citep[see e.g.][]{Patruno2021,Salvo2022}. Observations suggest that AMXPs were spun up in LMXBs through mass accretion during their earlier phases of long-term evolution. Furthermore, the discovery of transitional millisecond pulsars \citep[tMSPs;][]{archibald2009,papitto2013,bassa2014} strongly indicates that AMXPs eventually become radio millisecond pulsars (RMSPs), as predicted by the recycling scenario \citep{Alpar1982,Radhakrishnan1982}.

At present, there are 26 known AMXPs, including the three tMSPs. For all these sources, $P < 10$ ms \citep[for recent reviews, see][]{Patruno2021,Papitto2022}. For most AMXPs, the average spin period derivatives measured during the outbursts, $\Pdotoutburst$, were found to be negative \citep{Salvo2022}. For six AMXPs, the average $\Pdot$ during quiescent states, $\Pdotsecular$, estimated from $P$ variation, $\Delta P$, between the successive outbursts showed that these sources spin down during the quiescent states \citep{Salvo2022}. Measurements that include many outbursts and quiescent states also give positive average $\Pdot$ values, that is, they spin-down in the long-term. This long-term spin-down behaviour is likely to be common property of most of the other AMXPs as well considering their similar rotational rate and outburst behaviours. In the literature, AMXPs are usually assumed to be slowing down purely by the dipole torques during the quiescent states neglecting the disc--field interaction, due to low mass-inflow rate of the disc, $\Mdotin$. In this state, the inner disc could be evaporated by thermal instabilities \citep{Frank2002} for the sources with very low $\Mdotin$, and the magnetic dipole torque dominates the disc torque. For these sources, the dipole field strength at the NS equator is estimated from the dipole torque formula as $B \simeq 3.2\times 10^{19}~ \sqrt{P~\Pdot}$~G, using the measured $P$ and $\Pdotsecular$. If $\Mdotin$ levels do not decrease sufficiently for the evaporation of the inner disc during quiescence, the magnetic torque produced by the disc-field interaction dominates the dipole torque. In this case, the dipole torque formula overestimates the $B$ value.

The long-term rotational evolution of AMXPs is not easy to estimate due to their transient nature. Most of these sources are likely to enter the spin-up phase during the outbursts, while they spin-down in the quiescent state. The net torque calculated for a complete outburst--quiescence cycle depends on the model assumptions, and there is not a consensus on the magnitude and sign of the average disc torque acting on AMXPs. The long-term rotational evolution of a transient NS with an average disc mass-flow rate, $\langle \Mdotin\rangle$, is conventionally assumed to be the same as that of a persistent NS with a steady $\Mdotin$ that is equal to the $\langle \Mdotin\rangle$ of that transient source. Nevertheless, there are models that estimate much more efficient spin-up torques during the outbursts and weaker spin-down torques during quiescence of AMXPs such that the transient NS in the example above could have an equilibrium spin frequency much higher than that of the persistent source \citep[][hereafter BC2017]{DAngelo2017,Bhattacharyya2017}. In these models, AMXPs are estimated to reach sub-millisecond periods, and additional spin-down torque by gravitational radiation was suggested to explain the lack of sources with sub-millisecond periods (e.g. BC2017). In an alternative explanation for the lack of sources with sub-millisecond periods, it was shown that the correlation between the long-term accretion rate and the frozen dipole fields of LMXBs puts a natural barrier to the spin-up of LMXBs, preventing them from reaching the submillisecond periods \citep{ertan2021}.

\citet{Ertan2017,Ertan2018} proposed an analytical model to calculate the inner disc radius, $\rin$, and the torque--luminosity relation of NSs in the propeller phase. This model is based on the basic principles and the results of the simulations obtained with the model developed by \citet{Lovelace1995,Lovelace_1999} and \citet{Ustyugova_2006}. A detailed discussion can be found in \citet{Ertan2017}. The model was later developed to extend these calculations to include all the rotational phases and the transitions between these phases of NSs in LMXBs and to account for the typical torque--luminosity and torque reversal properties of NSs \citep{Ertan2020} with an application to the torque reversals of 4U 1626--67 \citep{Genali2022}. In this model, the $\rin$ value is usually estimated to be much smaller than in the conventional models for both the strong propeller and the spin-up phases. The extension of the inner disc to the NS surface can explain the lack of X-ray pulsations from most LMXBs \citep{Niang2024}. Motivated by these results, we use the same model to investigate the torques acting on AMXPs during their quiescence-outburst cycles. Our results imply that the spin-down torques due to magnetic dipole radiation and the disc-field interaction, and spin-up torques yielded by accretion on to the star are sufficient to account for the observed long-term spin-down of AMXPs without invoking additional spin-down torque mechanisms. We briefly describe the model in Section~\ref{model}. In Section~\ref{sources}, we present the properties of sources used in this study along with the model results obtained for each source. We discuss these results in Section~\ref{disccus} and summarize our conclusions in Section~\ref{conc}.

\section{The Model}\label{model}

 Here we briefly describe the model for the NSs accreting from geometrically thin accretion discs \citep[for details, see][]{Ertan2020}. In the model, there are three rotational phases: strong-propeller (SP), weak-propeller (WP), and spin-up (SU). These phases are determined by the location of $\rin$ with respect to $\rco$. There is not a single $\rin$ formula for all the rotational phases. We will describe the changing properties and phases of an illustrative model source with gradually increasing $\Mdotin$. 
 
 At low $\Mdotin$ levels, the system is in the SP phase. In this phase, the maximum inner disc radius at which the SP condition can be achieved is estimated as
 \begin{equation}        
        \Rinmax^{25/8}~|1-\Rinmax^{-3/2}| \simeq 0.22~\alpha_{-1}^{2/5}~ M_{1.4}^{-7/6}~\dot{M}_\mathrm{{in,16}}^{-7/20}~\umu_{26}~P_{-3}^{-13/12}
\label{eq:Rin}  
\end{equation}
where $\Rinmax = \rinmax/\rco$, $\alpha_{-1} = (\alpha/0.1)$ is the kinematic viscosity parameter \citep{Shakura1973}, $M_{1.4} = (M/1.4~\Msun)$ is the mass of the NS, $\dot{M}_\mathrm{{in,16}} = \Mdotin/(10^{16} ~\mathrm{g~s^{-1})}$, $\umu_{26}=\umu/(10^{26}\mathrm{~G~cm^3})$ is the magnetic dipole moment, and $P_{-3}$ is the spin period of the star in milliseconds. In this phase, $\rin =\reta = \eta~\rinmax$ where $\eta \lesssim 1$. We also define the radii $\Reta=\eta \Rinmax$, $R_{\ast}=r_{\ast}/\rco$ and $\Rxi=\xi \RA=\xi \rA /\rco=\rxi/\rco$ where $\rA$ is the conventional \Alfven radius, $\xi$ is a parameter close to unity, and $r_{\ast}\simeq10^6$~cm is the radius of the NS. In a steady SP phase, $\rin > r_1 = 1.26~\rco$ (point B in Fig.~\ref{fig:1751}), and all the inflowing disc matter is thrown out from the narrow inner disc boundary with speeds greater than the escape speed, $v_\mathrm{{esc}}$. In this phase, there is no mass accretion on to the NS, and the sources that have sufficient rotational power can emit ordinary radio pulses. 

\begin{table*}
\small
\centering
\caption{Properties of AMXPs. The $\Pdotoutburst$ values are obtained for the given pulsed $\Lx$ intervals.}
\label{tab1}
\renewcommand{\arraystretch}{1.4}
\begin{tabularx}{\textwidth}{
     >{\centering\arraybackslash}m{2.4cm}
     >{\centering\arraybackslash}m{1.0cm}
     >{\centering\arraybackslash}m{2.2cm}
     >{\centering\arraybackslash}m{2.1cm}
     >{\centering\arraybackslash}m{2.4cm}
     >{\centering\arraybackslash}m{2.3cm}
     >{\centering\arraybackslash}m{1.0cm} 
     >{\centering\arraybackslash}m{0.9cm} 
    }
\hline
AMXP Name & $P$ & $\Pdotoutburst$ & $\dot{P}_\mathrm{secular}$ & Pulsed $L_\mathrm{X}$ Interval & $L_\mathrm{X}$ (Quiescence) & Distance & Ref. \\
 & (ms) & ($10^{-18}$ s s$^{-1}$) & ($10^{-20}$ s s$^{-1}$) & ($10^{36}$ erg s$^{-1}$) & ($10^{32}$ erg s$^{-1}$) & (kpc) & \\
\hline
XTE J1751$-$305 & 2.3 & $-(1.96\pm 0.53)$ & $2.9\pm 0.6$ & $2.7-27.0$ & $< 4.6$ & 8.5& \textsuperscript{$a$}\\
IGR J17494$-$3030 & 2.66 & $> -12.7$ & $14.5\pm0.5$ &$ 0.05-2.2$ & $< 10.4$ & 8 & \textsuperscript{$b$} \\
Swift J1756.9$-$2508 & 5.5 & $>-9.05$ & $2.2\pm0.78$ & $3.0-9.6$ & $<10.0$ & 8 & \textsuperscript{$c$} \\
IGR J17511$-$3057 & 4.1 & $-(2.42 \pm 0.3)$ & $3.83\pm1.83$ & $3.0-11.0$ & $3.5-5.2$ & $<6.9$ & \textsuperscript{$d$}\\
IGR J00291$+$5934 & 1.7 & $-(1.42\pm 0.08)$ & $0.89$ & $0.3-8.1$ & $1.0-4.2$ & $4.2\pm0.5$ & \textsuperscript{$e$} \\
\hline
\end{tabularx}
\vskip 0.5em
\parbox{\textwidth}{
\small \textit{Note.} \textit{(a)} \citet{Markwardt2002,Wijnands2005,Gierliski2005,Papitto2008,Riggio2011a}; \textit{(b)} \citet{ArmasPadilla2013,Chakrabarty2013,Ng2020,Ng2021}; \textit{(c)} \citet{Krimm2007,Papitto2007,patruno2010b,Mukherjee2015,Bult2018}; \textit{(d)} \citet{Markwardt2009,Altamirano2010,Riggio2011b,sanna2025,Illiano2025}; \textit{(e)} \citet{Markwardt2004,Jonker2005,Galloway2005,Papitto2011,DeFalco2017}.
}
\end{table*}

As $\Mdotin$ increases, if $\rin$ is instantaneously between $r_1$ and $\rco$, the matter thrown out of the inner disc boundary falls back to the disc at larger radii. This causes a pile-up of the matter at the inner disc regions, which pushes the inner disc inwards until $\rin = \rco$ (from point B to D). This initiates the accretion on to the NS taking the system into the WP phase. In this phase, for a large range of $\Mdotin$ (from point D to E), the system spins down with mass flow from $\rin=\rco$ along the closed field lines on to the NS, which is likely to quench the ordinary radio pulses.

With further increase in $\Mdotin$, the viscous stresses dominate the magnetic stresses, and the inner disc penetrates into $\rco$. This corresponds to the WP/SU transition, which happens approximately when $\rxi\simeq\rco$. Beyond the $\Mdotin$ level for this transition, $\rin$ tracks $\rxi$ for a narrow $\Mdotin$ range until it reaches the unstable upper branch of $\reta$ (equation~(\ref{eq:Rin}); from point E to F). With $\Mdotin$ at point F, the inner disc moves inward, opening up the closed field lines until $\rin=\reta$ on the stable lower branch of $\reta$ \citep[see point G in fig. 1 of][]{Niang2024}. As $\Mdotin$ increases further, $\rin$ tracks $\reta$ until the disc reaches the surface of the NS \citep[see point H in fig. 1 of][]{Niang2024}. In some cases, stable $\reta$ corresponding to current $\Mdotin$ could be smaller than $r_\ast$, and the inner disc extends down to the NS surface (point H). This is the case for the five sources investigated in this work (see Figs~\ref{fig:1751}--\ref{fig:5934}).

The total torque acting on the star can be written as: 
\begin{equation}        
        \Gamma = \Mdot_\ast\sqrt{GM\rin} -\frac{\umu^2}{\rin^3 }\Bigg(\frac{\Delta r}{\rin}\Bigg)-\frac{2\umu^2\Omega^3_\ast}{3c^3}
\label{eq:torque}  
\end{equation}
where the first term is the spin-up torque, $\Gammaacc$, associated with the accretion on to the star. Here $G$ is the gravitational constant and $\Mdot_\ast$ is the mass accretion rate on to the NS. The second term is the spin-down torque, $\GammaD$, produced by the interaction between the magnetic dipole field lines and the disc inside the boundary region with radial width $\Delta r <r$. The last term is the magnetic dipole torque, $\Gammadip$, where $c$ is the speed of light and $\Omega_\ast$ is the angular frequency of the star. $\Gammadip$ is mostly negligible in the presence of $\GammaD$ and $\Gammaacc$.

In the SP phase, $\Mdotstar=0$ thus $\Gammaacc =0$, and the star slows down with $\Gamma=\GammaD+\Gammadip$. In the WP phase, $\Mdot_\ast=\Mdotin$ and all the torques are active, while $\GammaD$ mostly dominates both $\Gammaacc$ and $\Gammadip$, and the system slows down. In this phase, $\GammaD$ remains constant while $\Gammaacc$ increases with increasing $\Mdotin$ which eventually dominates $\GammaD$, causing the WP/SU transition (torque reversal). In the SU phase, $\Mdot_\ast=\Mdotin$ and $\Gamma=\Gammaacc+\Gammadip$.
 
We estimate $\Mdotstar$ from the X-ray luminosity, $\Lx =\Lacc=GM\Mdotstar/r_\ast$ produced by accretion on to the NS in the WP phase. In the SP phase, $\Mdotstar=0$ and $\Lx =L_{\mathrm{disc}}\simeq GM\Mdotin/2\rin$ emitted mostly from the inner disc. In the quiescent state, the sources are likely to be in either the SP phase or the WP phase in the model. It is also possible that the inner disc is truncated due to evaporation by thermal instabilities at low $\Mdotin$ levels in the quiescent state \citep{Frank2002}, and the source slows down with the magnetic dipole torque alone.

\section{The Sources and Results}\label{sources}

We analyse the five AMXPs with observational properties given in Table~\ref{tab1}. We excluded SAX J1808.4$-$3658, due to strong timing noise during its outbursts \citep[see e.g.][for a review]{Burderi2006,Salvo2022}. Among these AMXPs, the $\Pdotoutburst$ values were also measured for three sources and constrained for two sources (see Table~\ref{tab1}), while $\Pdotsecular$ were measured for the five sources. In our calculations, for comparison of a measured $\Pdotoutburst$ with the model result, we consider the $\Lx$ (and corresponding $\Mdotin$) range for which this particular $\Pdotoutburst$ was measured. Most of these sources enter the SU phase during their outburst states. For most AMXPs, there are discrepancies of a factor of a few between the measured $\Pdotoutburst$ values and those inferred from the conventional spin-up torque using the $\Mdotstar$ values estimated from the observed $\Lx$ range with detected X-ray pulsations during outbursts \citep[see e.g.][]{Burderi2007,Papitto2008,Salvo2022,Salvo2024}. In some cases, this discrepancy can be even more significant, as in the case of IGR J00291+5934 \citep{Burderi2007}. The torque magnitudes calculated in our model for the SU phase are similar to those estimated in the conventional models (see Fig.~\ref{fig:bha2}). Therefore, an agreement within a factor of a few in $\Pdotoutburst$ is sufficient for the purpose of this work.

For a given source, there are three different possibilities in the quiescent state: (1) the NS is in the SP phase with no accretion on to the star, (2) the system is in the WP phase with ongoing mass accretion on to the NS, or (3) the inner disc is truncated by evaporation and the NS is slowing down purely by the magnetic dipole torques. For each source studied here, we test all these three possibilities together with the outburst properties in the frame of our model as follows. For the cases (1) and (2), we calculated the $B$ values from our torque model, corresponding to the measured $\Pdotsecular$ values. For the case (3), $B$ that can produce observed $\Pdotsecular$ is calculated from the dipole-torque formula. The next step for each source is to estimate the $\Pdotoutburst$ in our model using the $B$ values estimated in the previous step, for each of the three cases, and compare our results with the observed $\Pdotoutburst$. The model curves together with observed data for the five sources are given in Figs~\ref{fig:1751}--\ref{fig:5934}. In all these figures, panels a, b, and c show the model curves obtained for the cases (1), (2), and (3) described above, respectively. With few exceptions our results mostly do not constrain the rotational phases of these sources in their quiescent states. The possibility of the evaporation of the inner disc in quiescence cannot be eliminated for some sources either, especially for the sources that have no $\Lx$ measurements during this state. Measuring $\Lx$ and $\Pdot$ of the sources simultaneously during quiescence and outburst would provide a better test for the models. We discuss the details of our results for each source below. For these different cases in quiescence, the model can produce reasonable results in agreement with the observed source properties in the outburst states. This implies that independent of their actual phases in quiescence, the long-term spin-down behaviour of these sources can be accounted for by $\GammaD$, $\Gammaacc$, and $\Gammadip$ without any need to an additional external spin-down torque. 

\subsection{XTE J1751--305}

\begin{figure*}
    \centering
    \includegraphics[width=\linewidth]{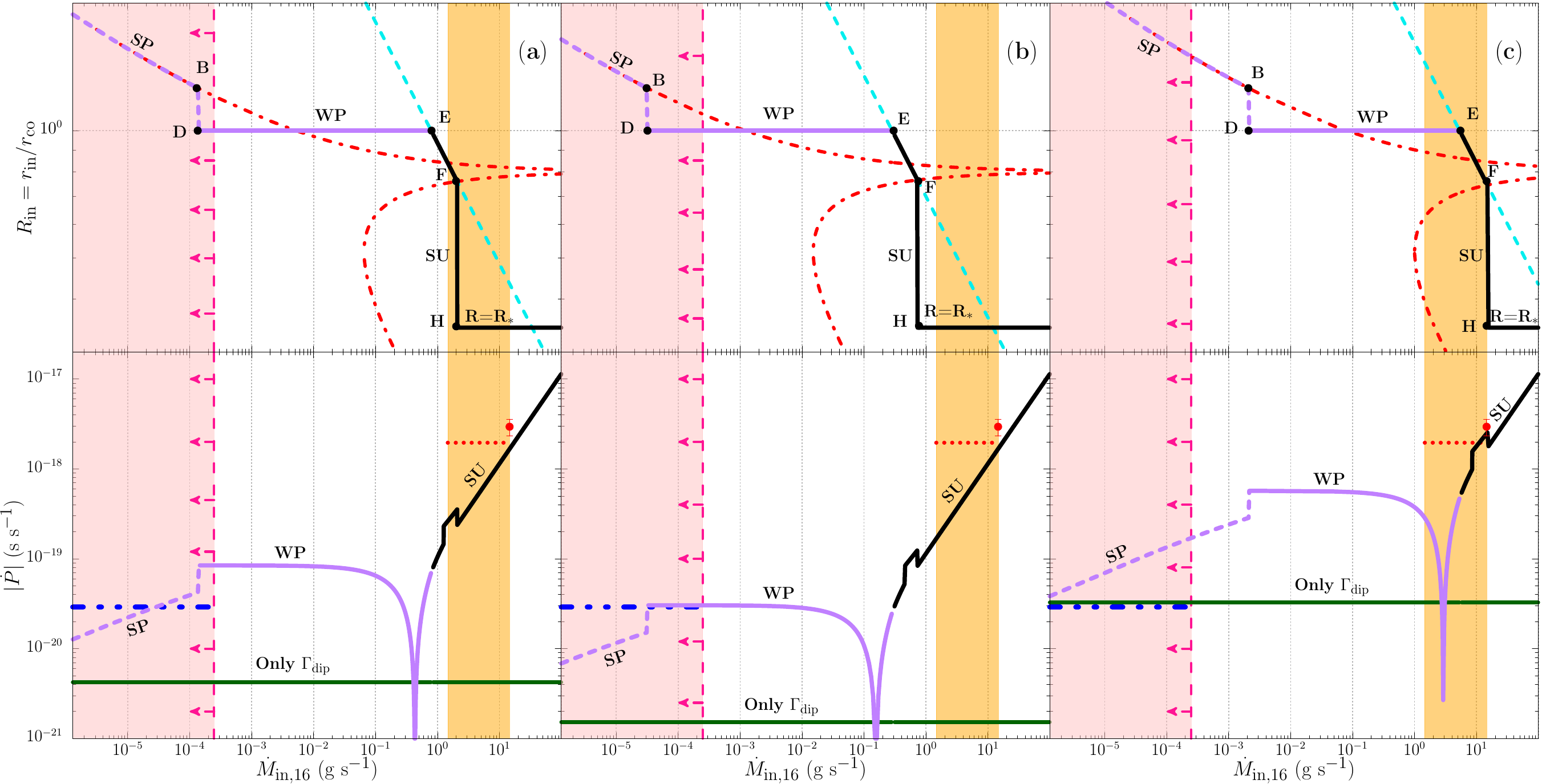}
   \caption{The model curves for XTE J1751$-$305. (a) The variation of $\Rin$ (upper panel) and $\Pdot$ (lower panel) with $\Mdotin$ across the SP (dashed purple curve), WP (solid purple curve), and SU (solid black curve) phases. All these model curves are obtained with $P = 2.3$ ms, $\Delta r/r_\mathrm{{in}} = 0.25$, $\eta = 0.8$, and $\xi = 0.8$. In the top panel, red dot-dashed curve shows the $\Reta$ solution and the turquoise dashed line represents the $\Rxi$. In the bottom panel, red dotted and blue dot-dot-dashed line segments denote the observed $\Pdotoutburst$ (negative) and $\Pdotsecular$ (positive) values, respectively. The vertical dashed line with the arrows corresponds to the upper limit on the quiescent $\Mdotin$ level. The orange shaded area in the SU phase shows the X-ray pulsed $\Mdotin$ range observed during the outburst. The red dot with error bars shows the measured $\Pdot$ value at the peak of the 2002 outburst. These model curves are obtained with $B\simeq 1.0\times 10^{8}$~G. (b) The same as (a) except $B\simeq 6.0\times10^{7}$~G. (c) The same as (a) except $B\simeq 2.6\times10^{8}$~G. The solid green lines show the $\Pdot$ values produced by the $\Gammadip$ alone. Our results indicate that all the three cases are possible for this source (see the text for the explanation).}
    \label{fig:1751}
\end{figure*}

This source was discovered in 2002 with $P = 2.3$ ms \citep{Markwardt2002}. The distance to the source is estimated to be $d\simeq8.5$~kpc assuming it is located near the Galactic Centre \citep{Gierliski2005}. Its observed properties and the model curves are seen in Fig.~\ref{fig:1751}. Coherent X-ray pulsations were detected during the first 9 days of the 2002 outburst \citep{Papitto2008} during which $\Lx$ decreased from the peak level of $2.7\times 10^{37}$~\ergpers~to $2.7\times10^{36}$~\ergpers~\citep[][orange shaded area in the SU region]{Gierliski2005}. For this $\Lx$ range, $\Pdotoutburst = -(1.96\pm 0.53)\times10^{-18}$~\spers~\citep[][horizontal dotted line segment]{Papitto2008} resulting in a spin-up of $\Delta P \simeq- 1.7\times10^{-12}$~s. The measurements close to the $\Lx$ peak of the 2002 outburst give $\Pdot=-(2.95\pm 0.63)\times10^{-18}$~\spers~\citep[][red data point with error bars]{Papitto2008}. Due to the low $\Lx$ levels and short durations in the 2005 and 2007 outbursts, the $P$ variations could not be measured. The $P$ measurements at the end of the 2002 outburst and at the beginning of the 2009 outburst give $\Pdotsecular = (2.90\pm 0.6)\times10^{-20}$~\spers~\citep[][ horizontal dot-dot-dashed line segment]{Riggio2011a} resulting in a spin-down of $\Delta P \simeq 6.4\times10^{-12}$~s. Comparing these $\Delta P$ values, it is seen that the spin-up during outbursts is not sufficient to change the long-term spin-down trend of the source. The $\Lx$ upper limit for the quiescent state is $4.6\times10^{32}$~\ergpers~\citep[][vertical dashed line with arrows]{Wijnands2005}.

It is seen in Fig.~\ref{fig:1751} that the source could be in the SP phase (Fig.~\figba) or possibly in the WP phase (Fig.~\figbb) during the quiescent state, depending on $\Mdotin$ and $B$. An illustrative model curve obtained with $B\simeq 1.0 \times10^8$~G in Fig.~\figba~reproduces the observed $\Pdotsecular$ with $\Mdotin\simeq2.85\times10^{11}$~\gpers~while the source is in the SP phase (no mass accretion). Since the actual $\Lx$ level during quiescence is unknown, the model cannot constrain the $B-\Mdotin$ pair that reproduces the observed $\Pdotsecular$. Lower $\Mdotin$ levels require higher $B$ values. Fig.~\figbb~shows the reasonable model curves if the source is in the WP phase (accretion allowed) in the quiescent state, which requires $B\simeq 6.0 \times10^7$~G to account for the observed $\Pdotsecular$. Note that for the models in Figs~\figba~and~\figbb, $\Gammadip$ (horizontal solid lines) is negligible compared to $\GammaD$. We also consider the possibility that the inner disc could be evaporated in the quiescent state (Fig.~\figbc). In this case, the spin-down with dipole torques can account for the measured $\Pdotsecular$ with $B\simeq 2.6 \times 10^{8}$~G. It is not easy to estimate the location of $\rin$ when the inner region of the disc is evaporated. We simply neglect $\GammaD$ in our calculations for the evaporated inner disc state. Independent of the inner disc properties in quiescence, the source enters the SU phase with sharply increasing $\Mdotin$ during the outburst. In Figs~\ref{fig:1751}--\ref{fig:5934}, orange shaded areas show the observed pulsed $\Mdotin$ ranges for the five sources. The $\Pdotoutburst$ measurements or upper limits obtained within these ranges are also denoted in the figures.

For XTE J1751$-$305, our results do not eliminate (or favour) the evaporation of the inner disc in quiescence. Detection of X-rays in the quiescent state could better constrain the models. For all three possibilities, the model can reproduce the observed $\Pdotoutburst$ and the $\Pdot$ values measured at the peak of the outburst. We note that the $\Pdot$ values in the model are estimated for sources with steady $\Mdotin$. Around the peak of the outburst, the inner disc condition could significantly deviate from the steady-state conditions. In particular, not only $\Mdotin$ but also $\rin$ varies sharply close to the peak of the outburst. In Fig.~\ref{fig:1751}, F to H transition requires extension of the inner disc towards the star in the viscous time-scale across the disc from $\sim\rco$ to $r_\ast$ during the rise phase of the outburst, which puts an uncertainty on the exact value of $\rin$ especially close to the peak and during the initial sharp decay of the outburst. Provided that the inner disc is close to the NS surface, the exact value of $\rin$ does not significantly affect the torque calculations. For our aim, it is sufficient to roughly produce the measured $\Pdot$ levels during outbursts.

\subsection{IGR J17494--3030}

\begin{figure*}
    \centering
    \includegraphics[width=\linewidth]{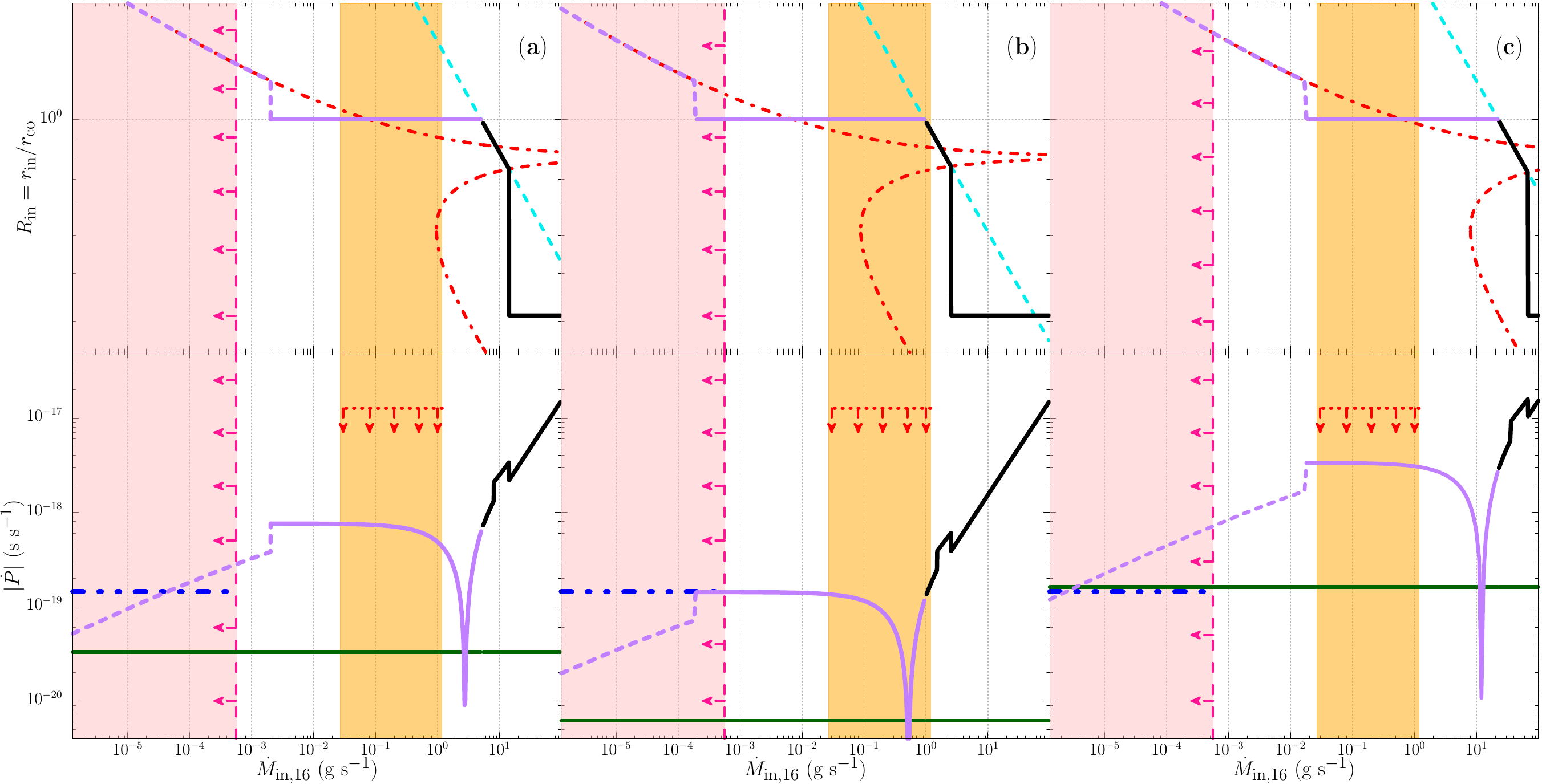}
   \caption{The model curves for IGR J17494$-$3030. The same as Fig.~\ref{fig:1751}, but with $P = 2.66$~ms. In bottom panels, the red arrows represents the upper limit on the $\Pdotoutburst$. These model curves are obtained with (a) $B \simeq 3.0\times 10^{8}$~G, (b) $B\simeq 1.3\times10^{8}$~G, and (c) $B\simeq 6.3\times10^{8}$~G (to test evaporated inner disc). Our results indicate that all the three cases are possible for this source (see the text for the explanation).}
    \label{fig:17494}
\end{figure*}

This source was discovered in 2012 \citep{Boissay2012,ArmasPadilla2013}. Assuming it is located close to the Galactic Centre, $d\simeq 8$~kpc \citep{ArmasPadilla2013}. The observed properties and the model curves for this source are seen in Fig.~\ref{fig:17494}. Two outbursts in 2012 and 2020 were observed with similar X-ray light curves \citep{Ng2021}. Coherent X-ray pulses with $P = 2.66$~ms were detected during the decay phase of the second outburst \citep{Ng2020}. The upper limit $|\Pdotoutburst|< 1.27\times 10^{-17}$~\spers~\citep[][horizontal dotted line segment with arrows]{Ng2021} was estimated for the $\Lx$ range from $2.2\times10^{36}$~\ergpers~to $5.0\times10^{34} $~\ergpers~\citep[][orange shaded area in the SU region]{Ng2021}. X-ray pulses were recovered from the reanalysis of the archived X-ray data of the 2002 outburst when $\Lx\simeq1.7\times10^{35}$~\ergpers~\citep{ArmasPadilla2013}, which allowed the estimation of $\Pdotsecular = {(1.45\pm0.05)\times 10^{-19}}$~\spers~\citep[][horizontal dot-dot-dashed line segment]{Ng2021}. Observed $\Pdot$ values yield spin-up upper limit $|\Delta P| < 8.8\times10^{-12}$~s during the $\sim$8-day outburst and spin-down $\Delta P \simeq 3.7\times10^{-11}$~s over $\sim$8 years of quiescence, indicating that the source is slowing down during its long-term evolution. For quiescence, $\Lx < 1.04\times10^{33}$~\ergpers~\citep[][vertical dashed line with arrows]{Chakrabarty2013}.

Our results from the model calculations can be summarised as follows. In the quiescent state, the source could be in either the SP phase with $B\simeq3.0 \times10^8$~G~(see Fig.~\figia) or the WP phase with $B\simeq1.3 \times10^8$~G~(see Fig.~\figib) remaining consistent with the measured $\Pdotsecular$ in both cases. During the entire outburst state, the source could be in the WP phase without any transition into the SP phase (see Fig.~\figia). Alternatively, depending on the value of $B$, it could be in the SU phase at the beginning of the outburst while $5.2\times10^{15}~\mathrm{g/s^{-1}} \lesssim\Mdotin\lesssim1.2\times10^{16}~\mathrm{g/s^{-1}}$, and enters into the WP phase for lower $\Mdotin$ values in the $5.2\times10^{15}
-2.7\times10^{14}~\mathrm{g/s^{-1}}$ range (see Fig.~\figib). In the case of evaporated inner disc in quiescence, the source slows down by the $\Gammadip$ with $B\simeq 6.3\times 10^{8}$~G (see Fig.~\figic). The inner disc forms again with a sharp increase in $\Mdotin$ taking the source into the WP phase in the outburst state similar to the outburst for the case seen in Fig.~\figia.

To sum up, our results do not constrain the state of the inner disc in quiescence. The model sources in the WP phase, SP phase, or evaporated disc states in quiescence are in agreement with the observed rotational properties of the source with reasonable $B$ values for both the quiescent and the outburst states.

\subsection{Swift J1756.9--2508}

\begin{figure*}
    \centering
    \includegraphics[width=\linewidth]{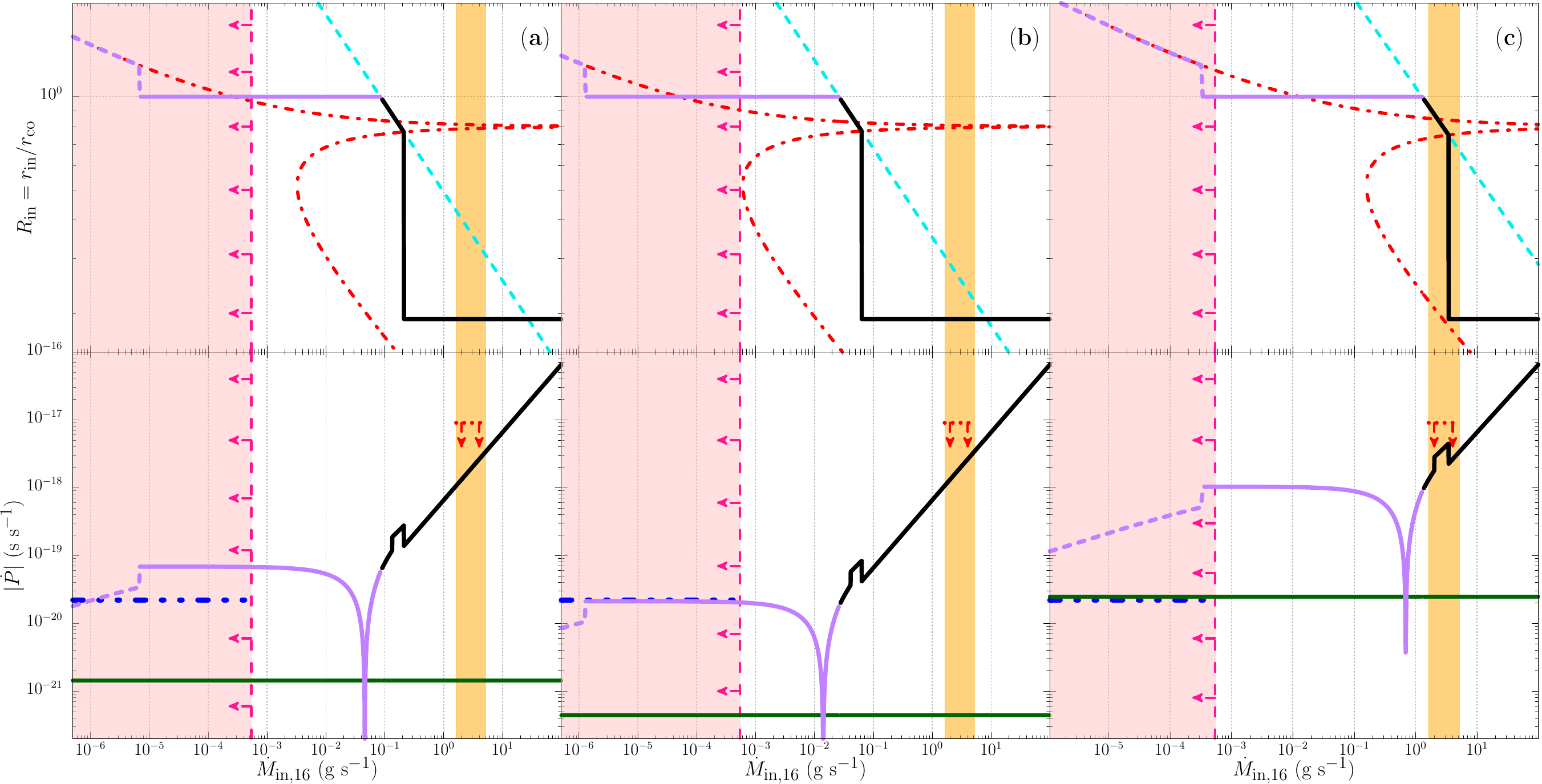}
   \caption{The model curves for Swift J1756.9$-$2508. The same as Fig.~\ref{fig:17494}, but with $P = 5.5$~ms. These model curves are obtained with (a) $B \simeq 9.0\times 10^{7}$~G, (b) $B\simeq 5.0\times10^{7}$~G, and (c) $B\simeq 3.5\times10^{8}$~G (to test inner disc evaporation). Our results indicate that all the three cases are possible for this source (see the text for the explanation).}
    \label{fig:1756}
\end{figure*}

This AMXP was discovered during the 2007 outburst with $P = 5.5$~ms \citep{Krimm2007}. This source is also assumed to be located near the Galactic Centre with $d\simeq 8$~kpc \citep{Krimm2007}. The source properties and the model curves are seen in Fig.~\ref{fig:1756}. The X-ray outburst light curves of 2009, 2018, and 2019 outbursts have peak $\Lx$ levels and durations similar to those of the 2007 outburst \citep{Patruno10,Sanna2018,Li2021}. The $P$ values were measured during all these outbursts. The upper limits on the $\Pdotoutburst$ were also obtained except for the 2019 outburst. In this work, we use the lowest upper limit obtained for the 2009 outburst, $|\Pdotoutburst| < 9.05\times10^{-18}$~\spers~\citep[][horizontal dotted line segment with arrows]{patruno2010b}. For the $\sim$12 days long 2009 outburst, $|\Delta P| < 9.4\times10^{-12}$~s. During this outburst, X-ray pulses were detected in the $\Lx$ range of $\simeq(3.0-9.6)\times10^{36}$~\ergpers~\citep[][orange shaded area in the SU region]{Mukherjee2015}. $\Pdotsecular=(2.2\pm 0.78)\times10^{-20}$~\spers~was estimated from the $P$ measurements during the 2007, 2009, and 2018 outbursts \citep[][horizontal dot-dot-dashed line segment]{Bult2018}, results in $\Delta P \simeq 7.6\times10^{-12}$~s. It is clearly shown in fig. 6 in \citet{Sanna2018} that the $\sim$2000 day measurements during the 2007, 2009, and 2018 outbursts indicate a net long-term spin-down. The upper limit on $\Lx$ during quiescence was found to be $\sim 1.0\times10^{33}$~\ergpers~\citep[][vertical dashed line with arrows]{Papitto2007}.

The model can reproduce the observed properties of Swift J1756.9--2508 with or without evaporation of the inner disc in quiescence. If the quiescent $\Lx$ level is close to the upper limit ($\sim10^{33}$~\ergpers), the source is likely to be in the WP phase with $\Lx=\Lacc$. For $\Lx\lesssim10^{31}$~\ergpers~($\Mdotin\lesssim10^{11}$~\gpers), the source could either be in the SP phase or have an evaporated inner disc, and slows down by the dipole torques. The field strengths estimated for the three cases in quiescence (WP phase, SP phase, or evaporated disc) are in the range of $5.0\times10^7-3.0\times10^8$~G~(see Fig.~\ref{fig:1756}). For all these cases, the sources are estimated to be in the SU phase for the observed $\Lx$ range during the outburst, with $\Pdot$ levels in agreement with the $\Pdotoutburst$ upper limit. 

\begin{figure*}
    \centering
    \includegraphics[width=\linewidth]{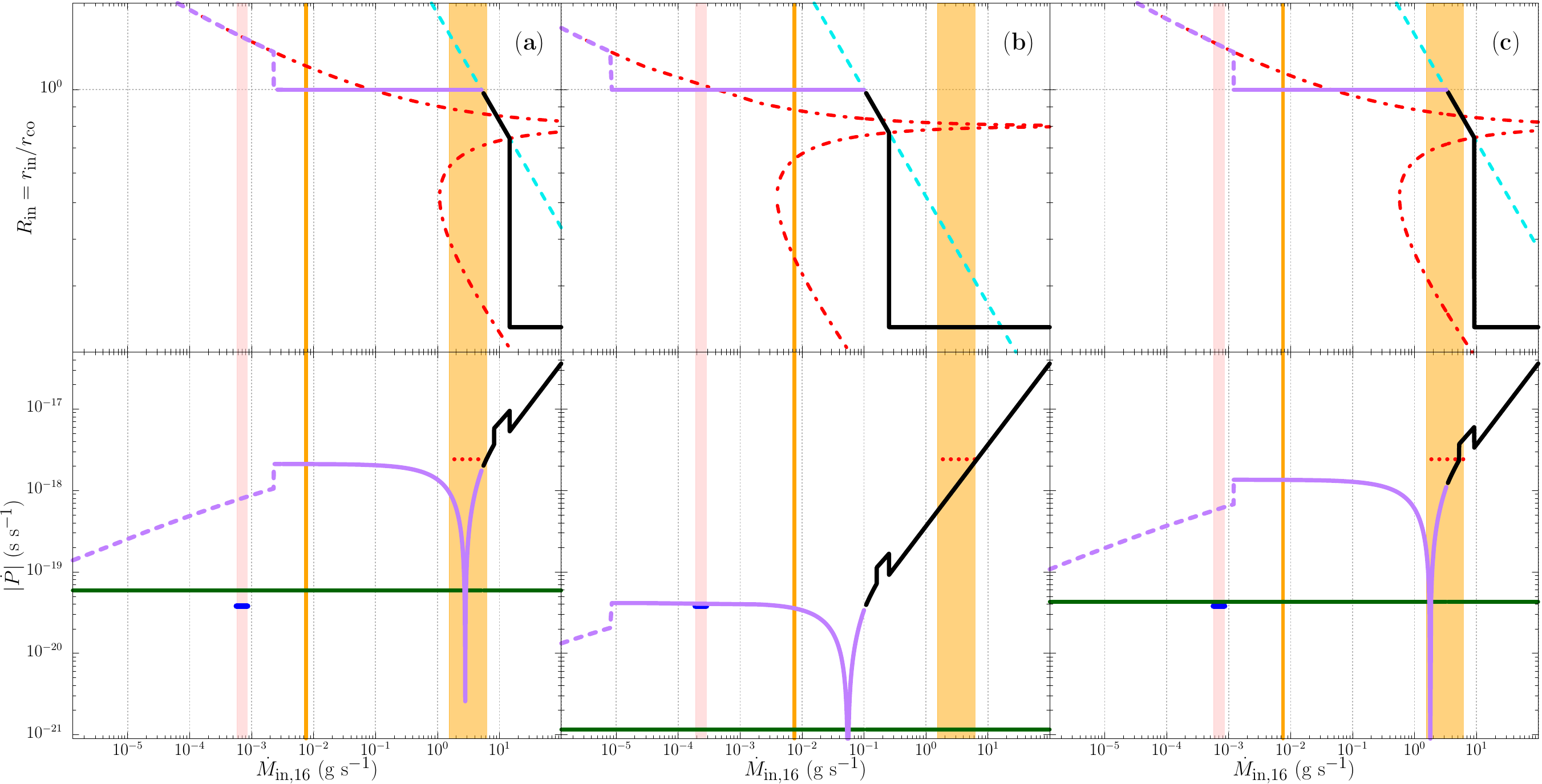}
   \caption{The model curves for IGR J17511$-$3057. The same as Fig.~\ref{fig:1751}, but with $P = 4.1$~ms. The orange vertical line represents the $\Lx$ value at which pulsations were observed. These model curves are obtained with (a) $B \simeq 5.0\times 10^{8}$~G, (b) $B\simeq 7.0\times10^{7}$~G, (c) $B\simeq 4.0\times10^{8}$~G (to test inner disc evaporation). Our results indicate that the cases (a) and (c) are not viable (see the text for the explanation).}
    \label{fig:3057}
\end{figure*}

\subsection{IGR J17511--3057}

This source was first detected with $P=4.1$~ms during the 2009 outburst \citep{Baldovin2009,Markwardt2009}. Similar subsequent outbursts were observed in 2015 and 2025 \citep{Bozzo2015,Sguera2025}. An upper limit of 6.9~kpc on the distance to the source was estimated from the type-I X-ray bursts observed during the 2009 outburst \citep{Altamirano2010}. Its observed properties and the model curves are seen in Fig.~\ref{fig:3057}. X-ray pulses were detected in all three outbursts, while $\Pdot$ was measured only during the 2009 outburst as $\Pdotoutburst=-(2.42 \pm 0.3) \times10^{-18}$~\spers~(horizontal dotted line segment) during which $\Lx$ decreased approximately from $1.1\times10^{37}$~\ergpers~to $3.0\times10^{36}$~\ergpers~\citep[][orange shaded area in the SU region]{Riggio2011b}. During this $\sim$23-day outburst, $P$ decreased by $\Delta P \simeq-4.8\times10^{-12}$~s. Pulsations were detected approximately 21 days after the onset of the 2025 outburst, when $\Lx\simeq1.4\times10^{34}$~\ergpers~was still $\sim20$ times greater than the quiescent level \citep[][orange vertical line]{Illiano2025}. From the $P$ measurements during the 2009, 2015, and 2025, it was estimated that $\Pdotsecular=(3.83\pm1.83)\times10^{-20}$~\spers~\citep[][horizontal blue line
segment]{sanna2025} with $\Delta P\simeq7.3\times10^{-12}$~s which exceeds $|\Delta P|$ during the spin-up (outburst) phase indicating a long-term spin-down for this source as well. For the quiescent state, $\Lx\sim(3.5-5.2)\times10^{32}$~\ergpers~was estimated from the archival Chandra observations in 2009 \citep[][pink strip in Fig.~\ref{fig:3057}]{Illiano2025}.

From the model calculations we obtained the following. This source is not likely to be in the SP phase during the quiescent state, since the $\Pdotsecular$ value estimated from the model exceeds the measured $\Pdotsecular$ by an order of magnitude (see Fig.~\figda). The source properties in quiescence can be produced in the WP phase with $B\simeq7.0\times10^7$~G (Fig.~\figdb), which is also consistent with the observed $\Pdotoutburst$ in the outburst state. During the outburst, the source enters the SU phase with abrupt rise in $\Mdotin$. For the case of evaporated disc in this state, illustrative model curves seen in Fig.~\figdc~are generated with $B\simeq4.0\times10^8$~G, consistently with the measured $\Pdotsecular$. However, $\Lx\sim10^{32}$~\ergpers~estimated for this state is not likely to be produced from the inner regions of a truncated disc, since it requires $\Mdotin>10^{13}$~\gpers, too high for the inner disc to be optically thin and thus truncated. Note that $\rin$ for a truncated disc should be greater than that estimated in our model, which requires even higher $\Mdotin$ to account for the observed $\Lx$ level. Recently, X-ray pulsations were detected at a low $\Lx$ level corresponding to $\Mdotin\simeq7.5\times10^{13}$~\gpers~\citep{Illiano2025} for which the source is estimated to be in the WP phase in the model which is consistent with the presence of X-ray pulsations.

\begin{figure*}
    \centering
    \includegraphics[width=\linewidth]{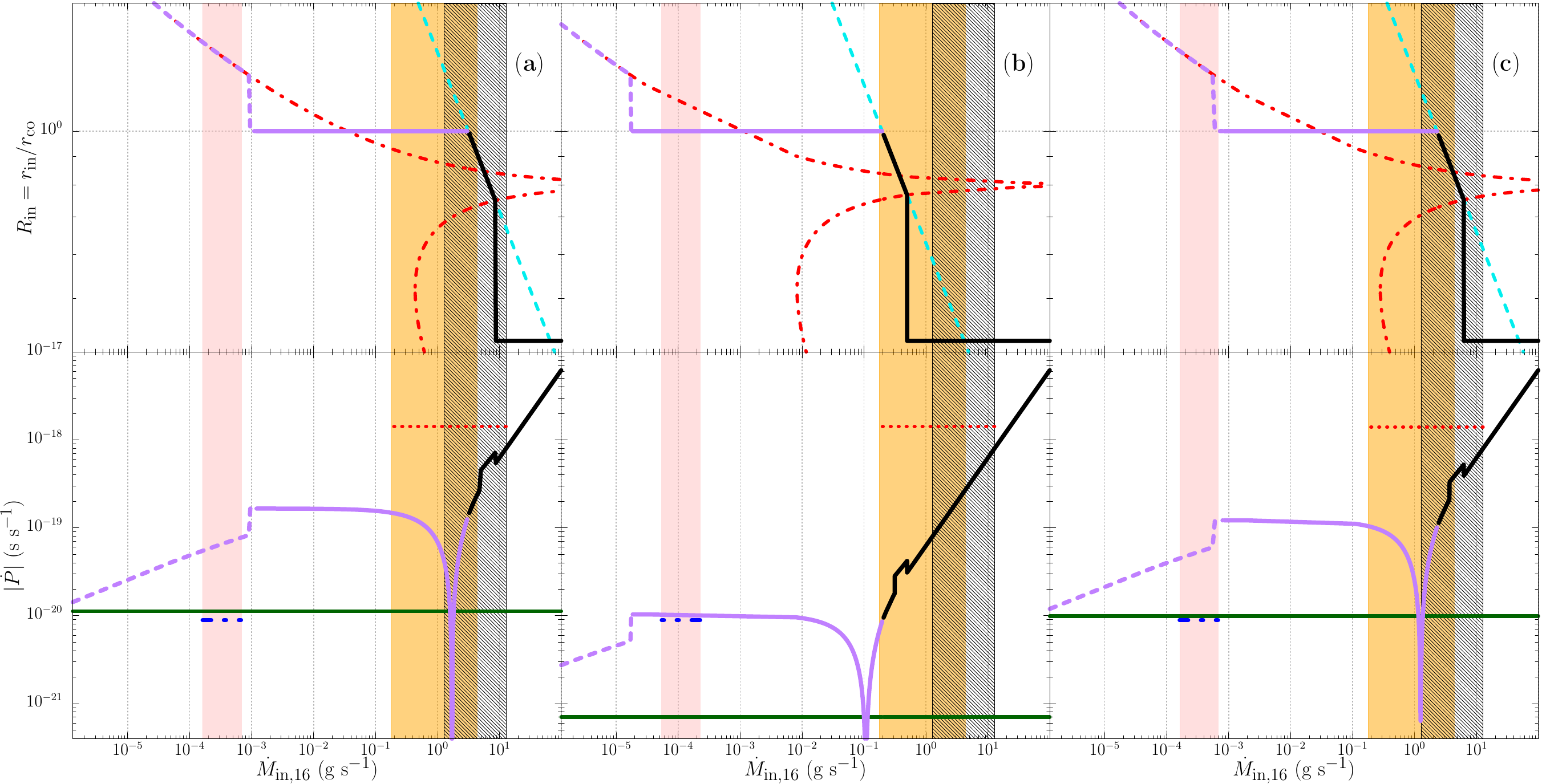}
   \caption{The model curves for IGR J00291$+$5934. The same as Fig.~\ref{fig:1751}, but with $P =1.7$~ms and the hatched $\Mdotin$ range corresponding to $L_\mathrm{X,peak}$ estimated with a different model (see text for details). These model curves are obtained with (a) $B\simeq 1.4\times 10^{8}$~G, (b) $B\simeq3.5\times10^{7}$~G, and (c) $B\simeq 1.2\times10^{8}$~G (to test inner disc evaporation). Our results indicate that the cases (a) and (c) are not possible (see the text for the explanation).}
    \label{fig:5934}
\end{figure*}

\subsection{IGR J00291+5934}\label{ıgr5934}

This source was discovered in 2004 as the fastest known AMXP with $P = 1.7$~ms \citep{Eckert2004,Markwardt2004}. The source exhibited three more X-ray outbursts; two in 2008 separated by~$\sim$30~days, each reaching roughly half the peak $\Lx$ of the 2004 outburst ($\sim4.0\times10^{36}$~\ergpers) with durations of 9 to 15 days and the third in 2015, which had the highest peak $\Lx$ ($\sim1.0\times10^{37}$~\ergpers) with a duration of 25 days \citep{Patruno2010,DeFalco2017}. The distance to the source was estimated to be $4.2\pm0.5$~kpc from PRE during the type-I X-ray burst observed in the 2015 outburst \citep{DeFalco2017}. Its observed properties and the model curves are seen in Fig.~\ref{fig:5934}. The $P$ and $\Pdotoutburst$ values were measured in all outbursts \citep{Patruno2010,Sanna2017}. In this work, we only use the $\Pdotoutburst$ of the 2004 outburst, since the 2008 outbursts were unusual and the timing analysis of the 2015 outburst did not cover the entire outburst \citep{Sanna2017}. 

For the timing of the 2004 outburst, RXTE PCA (PCU) data from 3–21 December 2004 are used in the published analyses \citep{Falanga2005,Galloway2005,Burderi2007, Patruno2010,Papitto2011}. These works report that pulsations became insignificant (rms $<1\%$) after December 14. \citet{Burderi2007} showed that different timing models infer different $\Pdotoutburst$ values ($-[(2.4\pm0.3)-(3.3\pm0.4)]\times10^{-18}~\mathrm{s~s^{-1}}$), and that the available statistics are not sufficient to decide on a particular timing model. \citet{Patruno2010} attempted to separate the effects due to flux-related phase variations by considering motion of the hot spot on the NS surface and reported a refined value, $\Pdotoutburst = -(1.42\pm0.08) \times 10^{-18}$~\spers, noting that this result is still subject to uncertainties related to timing noise and the adopted timing model. In an independent work, \citet{Papitto2011} also obtained the same $\Pdotoutburst$ value. For this source, chosen base-line also seems to affect $\Pdotoutburst$ measurement significantly \citep{Patruno2010}. For instance, splitting the data into two segments, December 3–10 ($\Pdotoutburst = -(1.0\pm0.1) \times 10^{-18}$~\spers) and December 10 to the end of the outburst ($\Pdotoutburst = -(6.7\pm0.1) \times 10^{-18}$~\spers), yields spin-up torque magnitudes increasing with decreasing X-ray flux, which is not expected in any spin-up torque model and also indicates significant timing noise effects on these analyses \citep{Patruno2010}. For the 2015 outburst, $\Pdotoutburst=-(8.4)\times10^{-18}$~\spers, approximately six times greater than the value in the 2004 outburst, while the 2004 and 2015 X-ray outburst light curves are similar \citep{Sanna2017}. In Fig.~\ref{fig:5934}, red dotted line segment in the bottom panels corresponds to $\Pdotoutburst = -(1.42\pm0.08) \times 10^{-18}$~\spers.

For the 2004 outburst, the pulsed $\Lx$ ($0.1-200$~keV) decreased from the peak value $8.0\times10^{36}$~\ergpers~to $ 3.0 \times10^{35}$~\ergpers~in approximately two weeks with a bolometric correction factor of 2.54 estimated from the broadband spectral fits \citep[][orange shaded area]{Galloway2005}. The most detailed analysis of the bolometric peak X-ray luminosity, $L_{\rm X,peak}$, was performed by \citet{Burderi2007} considering the beaming effects and contributions from the disc, which gives $L_{\rm X,peak} \simeq 2.5 \times 10^{37}$~\ergpers~(assuming a distance of 5 kpc). However, they concluded that a higher $\Mdotin$ than that inferred from the observed $\Lx$ is required to account for the measured $\Pdotoutburst$, corresponding to $L_{\rm X,peak} \sim 10^{38}$~\ergpers. The hatched area in Fig.~\ref{fig:5934} represents the $\Lx$ range with detected X-ray pulsations, estimated by \citet{Burderi2007} for $d=4.2\pm0.5$~kpc, decreasing from $\sim2.4\times10^{37}$~\ergpers~to $\sim2.4\times10^{36}$~\ergpers. These uncertainties in both $\Lx$ estimation and $\Pdot$ measurement of this source should be taken into account when comparing the observations with testing the model calculations.

During quiescence from the end of the 2004 outburst to the beginning of the first 2008 outburst, $\Pdotsecular= (1.14\pm 0.33) \times10^{-20}$~\spers~\citep{Papitto2011}. We have estimated $\dot{P}_\mathrm{secular} \simeq 8.9 \times 10^{-21}$~s~s$^{-1}$ (horizontal dot-dot-dashed line segment) using the $P$ values measured at the end of the 2004 and at the beginning of the 2015 outbursts \citep{Papitto2011,DeFalco2017}, assuming that the fainter outbursts with short duration in 2008 did not spin up the source significantly. This gives a spin-down of $\Delta P\simeq3.1\times10^{-12}$~s, which indicates that the source spins down on the average. In the quiescent state, $\Lx\sim(1.0-4.2)\times10^{32}$~\ergpers~\citep[][pink shaded area]{Jonker2005}.

From the model calculations, we find that the source is not likely to be in the SP phase in quiescence, as the measured $\Pdotsecular$ is almost an order of magnitude lower than the level estimated for this phase (see Fig.~\figbea). The measured $\Pdotsecular$ value can be achieved in the WP phase with $B\simeq3.5\times10^7$~G (Fig.~\figbeb). If the inner disc is evaporated in quiescence, the magnetic dipole torque requires $B\simeq1.2\times10^8$~G~to account for the observed $\Pdotsecular$ (Fig.~\figbec). However, as in the case of IGR J17511--3057, the high $\Mdotin$ during quiescence required to yield the observed $\Lx$ is not consistent with an optically thin and truncated inner disc. In other words, the disc is not likely to be evaporated during quiescence. Thus, the model favours the WP phase for this source during quiescence. In the outburst state, the source initially enters the SU phase and later transitions into the WP phase when $\Mdotin$ decreases below the critical level. With the $B$ value that can produce the measured $\Pdotsecular$ in the WP phase, the model cannot yield the measured $\Pdotoutburst$. For the lowest estimated $L_\mathrm{X,peak}$ values (orange shaded area) and the relatively high $L_\mathrm{X,peak}$ (hatched area) estimated by \citet{Burderi2007}, the $\Pdotoutburst$ levels estimated in our model remains below the measured value by factors $\sim5$ and $\sim2$ respectively (see Fig.~\ref{fig:5934}). In the model, $L_\mathrm{X,peak}\sim5\times10^{37}$~\ergpers~is required to remain consistent with the observed $\Pdotoutburst$ value. This discrepancy is reasonable considering the uncertainties in both the measured $\Lx$ and $\Pdotoutburst$ mentioned above and the simplified time-independent model calculation for the dynamical outburst state.

\section{Discussion}\label{disccus}

We have investigated the $\Lx$ and the rotational properties of the five AMXPs during their outbursts and quiescent states with the model used in earlier work to explain some typical torque-luminosity relations and the X-ray pulsation behaviours of LMXBs \citep{Ertan2020,Genali2022,Niang2024}. Our results indicate that the spin-down torques produced by disc-field interaction and magnetic dipole radiation, and the spin-up torque associated with accretion on to the NS are sufficient to explain the net torques acting on the NS during the outburst and quiescent states, which determine the long-term spin-down behaviours of most of these systems without requiring any additional spin-down torque. 

\begin{figure}
    \centering
    \includegraphics[width=\linewidth]{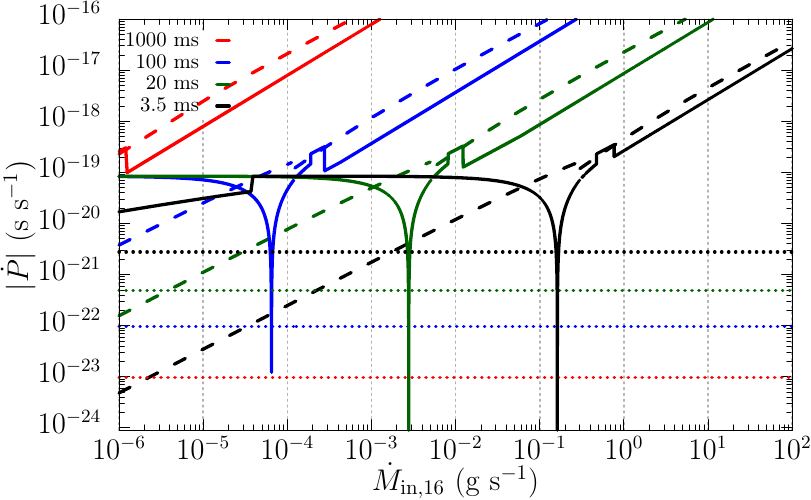}
   \caption{The model curves are produced with the same model parameters used in Figs~\ref{fig:1751}--\ref{fig:5934}, but with $B=1.0\times10^8$~G and different $P$ values (given in the figure). The solid curves are obtained with the analytical model described in Section~\ref{model}. The dotted horizontal lines show the $\Pdot$ values produced only by $\Gammadip$. The dashed curves are given by the torque formulas employed by BC2017.}
    \label{fig:bha2}
\end{figure}

In our model, the magnitude of the spin-down torque in the quiescent state is smaller and comparable to that of spin-up torque in the outburst states, while the duration of the quiescent phase is longer than that of the outburst phase (see Fig.~\ref{fig:bha2}). This naturally produces the average spin-down trend of these systems. Even if the spin-up torques are stronger in some other models (e.g. BC2017), more efficient torques in quiescence in our model are sufficient to account for the long-term spin down of these systems. We note that the torque reversals of LMXBs with well defined properties, in particular similar torque magnitudes on either side of the reversal which occurs with a small variation in $\Lx$, provide an important test for the torque models to be used to analyse the long-term evolution of these systems \citep{Bildsten1997,Takagi2016}.

In some conventional torque models, the spin-down torques produced by the disc-field interaction are much weaker than in our model, which necessitates additional spin-down torques to account for the observed long-term net torque acting on AMXPs. In these models, to explain the lack of sub-millisecond pulsars, gravitational radiation (GR) was suggested to be the external spin-down mechanism preventing these systems from reaching the sub-millisecond periods \citep[BC2017;][]{Bildsten_1998}. The GR torque could contribute to the long-term rotational evolution of AMXPs, particularly for the fastest rotators, since GR torque $\Gamma_{\mathrm{GR}} \propto P^{-5}$. In this context, \citet{Haskell_2011} suggested that spin equilibrium can be achieved through disc-magnetosphere interaction. It was shown by \citet{ertan2021} that the observed minimum $P$ of NSs can be explained by the natural barrier to the spin-up of NSs as a result of the correlation estimated between the long-term mass accretion rate and the frozen dipole field. Without ruling out possible contributions from the gravitational radiation to the angular momentum loss, several authors have suggested that the observed secular spin-down of even the fastest sources, such as IGR J00291$+$5934 and SAX J1808.4$-$3658, can be explained by the magnetic dipole radiation alone \citep[see e.g.][]{Hartman2008,Patruno2010,Papitto2011}. Considering both outburst and quiescence states we have found that $\Gammaacc$, $\GammaD$, and $\Gammadip$ produce rotational evolutions that are in agreement with the long-term spin-down behaviour of the sources. Due to uncertainty in X-ray luminosity and noise effects in timing, our results are not sufficient to rule out the spin-down contribution of GR either.

For a given $B$, model curves produced with different $P$ values are presented in Fig.~\ref{fig:bha2}. It is seen that the morphologies of the torque reversals are similar for sources with rather different critical $\Mdotin$ levels for torque reversals, and also in good agreement with the torque reversals of 4U 1626--67 \citep{Genali2022}. For comparison, dashed lines show the $|\Pdot|$ variation with $\Mdotin$ estimated in BC2017. It is seen that the spin-down torque magnitudes estimated in our model are orders of magnitude greater than in BC2017 for large ranges of $\Mdotin$. This is mainly due to the property of the WP phase with $\rin=\rco$, for which $\GammaD$ is constant and dominates $\Gammaacc$ when $\Mdotin$ is not close to the critical level for the torque reversal. For the SP phase, the dependence of $\rin$ on $\Mdotin$ in our model is much weaker than $\rA$, which also give relatively strong torques in the SP phase in comparison with the conventional models \citep[see][for details]{Ertan2020}.

Pulsed radio emission was observed from three tMSPs which show transitions between the X-ray pulsar and the radio pulsar states. The other AMXPs have not been observed to show radio pulses in the quiescent states. In our model, the inner disc radius, and thus the critical accretion rate for the WP/SP transition depends on $P$, unlike in conventional models, as well as $\Mdotin$ and $B$. The critical $\Mdotin$ for this transition predicted in \citet{Ertan2017,Ertan2018} is in agreement with the transition properties of tMSPs. Estimated critical $\Mdotin$ levels, which are not achieved in other AMXPs are discussed in \citet{Ertan2018}. 

Our results imply that the five sources investigated in this work could be in either the SP (no accretion) or WP (ongoing mass accretion) phase during the quiescent state if the inner disc is not evaporated. For IGR J17511--3057 and IGR J00291+5934, our model results exclude both SP phase and the evaporated disc in quiescence. For IGR J00291+5934, $\Pdotoutburst$ estimated with the model is a few times smaller than the observed level which has some uncertainties (see Section~\ref{ıgr5934}). The model does not eliminate the possibility that the inner disc could be evaporated for the sources that have upper limits on $\Lx$ in the quiescent state. If a given system is in the WP phase, the pulsed radio emission is hindered by the accretion on to the NS. The situation is not clear in the evaporated inner disc state. It could be the case that spherical like geometry of the hot matter evaporated from the inner disc may cause a fraction of this matter flow on to the poles of the star quenching the pulsed radio emission. Pulsed radio emission is allowed in the SP phase. Pulsed radio emission epochs of tMSPs correspond to SP phases in our model.

AMXPs are observed during the transient epochs of the long-term evolution of LMXBs. The results of a recent study, which uses the stellar evolution code MESA and our analytical model, show that the NS and binary properties of individual AMXPs can be reproduced simultaneously \citep{niang2026}. In an independent work, we will extend this approach to calculate the rotational properties of AMXPs during their outburst-quiescence cycles.

\section{Conclusion}\label{conc}

We have investigated the long-term evolution of AMXPs using the analytical torque model developed by \citet{Ertan2020}. For the five AMXPs analysed in this work, measured $\Pdotsecular$ and $\Pdotoutburst$ values show that these sources slow down on the average. During the quiescent state, if the inner disc is not evaporated by thermal instabilities, our results indicate that the sources could be in either the SP phase or the WP phase depending on the actual values of $B$, $\Mdotin$, and $P$. We have also tested the possibility of the inner disc evaporation due to thermal instabilities at low $\Mdotin$ levels in the quiescent state. In this case, the source evolves with magnetic dipole torques alone in quiescence. Our results do not eliminate this possibility for three sources. The model results constrain the phases of IGR J17511--3057 and IGR J00291+5934 in quiescence. The observed properties of these sources can be explained in the model only if they are in the WP phase in quiescence. The measured $\Pdotoutburst$ for IGR J00291+5934 can be reproduced consistently with the source properties in quiescence with an $\Lx$ level a few times greater than the highest $\Lx$ estimated by \citet{Burderi2007}, which seems to be reasonable considering timing noise and uncertainties in $\Lx$ estimation (see Section~\ref{ıgr5934} for details).

Our results estimated from the dipole and disc torques in our model are compatible with their observed long-term spin-down behaviour of the sources with reasonable model parameters. The basis of the relatively strong spin-down torques in the WP and SP phases in our model is the range of $\rin$ values which are estimated to be much smaller than in the conventional models. Our results do not exclude GR torques due to large uncertainties in both $\Lx$ and timing measurements. The cumulative spin-down during the long quiescent states dominates the spin-up by accretion during outbursts. In future work, we will investigate the long-term evolution of the transient LMXBs with different outburst and quiescent state properties.

\section*{Acknowledgements}

We acknowledge research support from Sabanc{\i} University, and from T\"{U}B\.{I}TAK (The Scientific and Technological Research Council of Turkey) through grant 123F083. We thank M. Ali Alpar for useful comments which significantly improved the manuscript.

\section*{Data Availability}
No new data were analysed in support of this paper.



\bibliographystyle{mnras}
\bibliography{export} 

\begin{thebibliography}{}
\makeatletter
\relax
\def\mn@urlcharsother{\let\do\@makeother \do\$\do\&\do\#\do\^\do\_\do\%\do\~}
\def\mn@doi{\begingroup\mn@urlcharsother \@ifnextchar [ {\mn@doi@}
  {\mn@doi@[]}}
\def\mn@doi@[#1]#2{\def\@tempa{#1}\ifx\@tempa\@empty \href
  {http://dx.doi.org/#2} {doi:#2}\else \href {http://dx.doi.org/#2} {#1}\fi
  \endgroup}
\def\mn@eprint#1#2{\mn@eprint@#1:#2::\@nil}
\def\mn@eprint@arXiv#1{\href {http://arxiv.org/abs/#1} {{\tt arXiv:#1}}}
\def\mn@eprint@dblp#1{\href {http://dblp.uni-trier.de/rec/bibtex/#1.xml}
  {dblp:#1}}
\def\mn@eprint@#1:#2:#3:#4\@nil{\def\@tempa {#1}\def\@tempb {#2}\def\@tempc
  {#3}\ifx \@tempc \@empty \let \@tempc \@tempb \let \@tempb \@tempa \fi \ifx
  \@tempb \@empty \def\@tempb {arXiv}\fi \@ifundefined
  {mn@eprint@\@tempb}{\@tempb:\@tempc}{\expandafter \expandafter \csname
  mn@eprint@\@tempb\endcsname \expandafter{\@tempc}}}

\bibitem[\protect\citeauthoryear{Alpar, Cheng, Ruderman  \& Shaham}{Alpar
  et~al.}{1982}]{Alpar1982}
Alpar M.~A.,  Cheng A.~F.,  Ruderman M.~A.,   Shaham J.,  1982, \mn@doi
  [Nature] {10.1038/300728A0}, 300, 728

\bibitem[\protect\citeauthoryear{Altamirano, Watts, Linares, Markwardt,
  Strohmayer  \& Patruno}{Altamirano et~al.}{2010}]{Altamirano2010}
Altamirano D.,  Watts A.,  Linares M.,  Markwardt C.~B.,  Strohmayer T.,
  Patruno A.,  2010, \mn@doi [Monthly Notices of the Royal Astronomical
  Society] {10.1111/j.1365-2966.2010.17369.x}, 409, 1136

\bibitem[\protect\citeauthoryear{{Archibald} et~al.,}{{Archibald}
  et~al.}{2009}]{archibald2009}
{Archibald} A.~M.,  et~al., 2009, \mn@doi [Science] {10.1126/science.1172740},
  \href {https://ui.adsabs.harvard.edu/abs/2009Sci...324.1411A} {324, 1411}

\bibitem[\protect\citeauthoryear{Armas~Padilla, Wijnands  \&
  Degenaar}{Armas~Padilla et~al.}{2013}]{ArmasPadilla2013}
Armas~Padilla M.,  Wijnands R.,   Degenaar N.,  2013, \mn@doi [Monthly Notices
  of the Royal Astronomical Society: Letters] {10.1093/mnrasl/slt119}, 436

\bibitem[\protect\citeauthoryear{Bahramian \& Degenaar}{Bahramian \&
  Degenaar}{2022}]{Bahramian2022}
Bahramian A.,  Degenaar N.,  2022, Low-Mass X-ray Binaries.
Springer Nature Singapore, Singapore, pp 1--62,
  \mn@doi{10.1007/978-981-16-4544-0_94-1}, \url
  {https://doi.org/10.1007/978-981-16-4544-0_94-1}

\bibitem[\protect\citeauthoryear{{Baldovin} et~al.,}{{Baldovin}
  et~al.}{2009}]{Baldovin2009}
{Baldovin} C.,  et~al., 2009, The Astronomer's Telegram, \href
  {https://ui.adsabs.harvard.edu/abs/2009ATel.2196....1B} {2196, 1}

\bibitem[\protect\citeauthoryear{{Bassa} et~al.,}{{Bassa}
  et~al.}{2014}]{bassa2014}
{Bassa} C.~G.,  et~al., 2014, \mn@doi [\mnras] {10.1093/mnras/stu708}, \href
  {https://ui.adsabs.harvard.edu/abs/2014MNRAS.441.1825B} {441, 1825}

\bibitem[\protect\citeauthoryear{Bhattacharyya \& Chakrabarty}{Bhattacharyya \&
  Chakrabarty}{2017}]{Bhattacharyya2017}
Bhattacharyya S.,  Chakrabarty D.,  2017, \mn@doi [The Astrophysical Journal]
  {10.3847/1538-4357/835/1/4}, 835, 4

\bibitem[\protect\citeauthoryear{Bildsten}{Bildsten}{1998}]{Bildsten_1998}
Bildsten L.,  1998, \mn@doi [The Astrophysical Journal] {10.1086/311440}, 501,
  L89

\bibitem[\protect\citeauthoryear{Bildsten et~al.,}{Bildsten
  et~al.}{1997}]{Bildsten1997}
Bildsten L.,  et~al., 1997, \mn@doi [The Astrophysical Journal Supplement
  Series] {10.1086/313060/XML}, 113, 367

\bibitem[\protect\citeauthoryear{{Boissay} et~al.,}{{Boissay}
  et~al.}{2012}]{Boissay2012}
{Boissay} R.,  et~al., 2012, The Astronomer's Telegram, \href
  {https://ui.adsabs.harvard.edu/abs/2012ATel.3984....1B} {3984, 1}

\bibitem[\protect\citeauthoryear{{Bozzo} et~al.,}{{Bozzo}
  et~al.}{2015}]{Bozzo2015}
{Bozzo} E.,  et~al., 2015, The Astronomer's Telegram, \href
  {https://ui.adsabs.harvard.edu/abs/2015ATel.7275....1B} {7275, 1}

\bibitem[\protect\citeauthoryear{Bult et~al.,}{Bult et~al.}{2018}]{Bult2018}
Bult P.,  et~al., 2018, \mn@doi [The Astrophysical Journal]
  {10.3847/1538-4357/aad5e5}, 864, 14

\bibitem[\protect\citeauthoryear{Burderi, Salvo, Menna, Riggio  \&
  Papitto}{Burderi et~al.}{2006}]{Burderi2006}
Burderi L.,  Salvo T.~D.,  Menna M.~T.,  Riggio A.,   Papitto A.,  2006,
  \mn@doi [The Astrophysical Journal] {10.1086/510666/FULLTEXT/}, 653, L133

\bibitem[\protect\citeauthoryear{Burderi et~al.,}{Burderi
  et~al.}{2007}]{Burderi2007}
Burderi L.,  et~al., 2007, \mn@doi [The Astrophysical Journal]
  {10.1086/510659}, 657, 961

\bibitem[\protect\citeauthoryear{{Chakrabarty}, {Jonker}  \&
  {Markwardt}}{{Chakrabarty} et~al.}{2013}]{Chakrabarty2013}
{Chakrabarty} D.,  {Jonker} P.~G.,   {Markwardt} C.~B.,  2013, The Astronomer's
  Telegram, \href {https://ui.adsabs.harvard.edu/abs/2013ATel.4886....1C}
  {4886, 1}

\bibitem[\protect\citeauthoryear{D'Angelo}{D'Angelo}{2017}]{DAngelo2017}
D'Angelo C.~R.,  2017, \mn@doi [Monthly Notices of the Royal Astronomical
  Society] {10.1093/MNRAS/STX1306}, 470, 3316

\bibitem[\protect\citeauthoryear{{De Falco}, {Kuiper}, {Bozzo}, {Galloway},
  {Poutanen}, {Ferrigno}, {Stella}  \& {Falanga}}{{De Falco}
  et~al.}{2017}]{DeFalco2017}
{De Falco} V.,  {Kuiper} L.,  {Bozzo} E.,  {Galloway} D.~K.,  {Poutanen} J.,
  {Ferrigno} C.,  {Stella} L.,   {Falanga} M.,  2017, \mn@doi [\aap]
  {10.1051/0004-6361/201629575}, \href
  {https://ui.adsabs.harvard.edu/abs/2017A&A...599A..88D} {599, A88}

\bibitem[\protect\citeauthoryear{Di~Salvo \& Sanna}{Di~Salvo \&
  Sanna}{2022}]{Salvo2022}
Di~Salvo T.,  Sanna A.,  2022, in Bhattacharyya S.,  Papitto A.,   Bhattacharya
  D.,  eds, , Millisecond Pulsars.
Springer International Publishing, Cham, pp 87--124,
  \mn@doi{10.1007/978-3-030-85198-9_4}, \url
  {https://doi.org/10.1007/978-3-030-85198-9_4}

\bibitem[\protect\citeauthoryear{Di~Salvo, Papitto, Marino, Iaria  \&
  Burderi}{Di~Salvo et~al.}{2024}]{Salvo2024}
Di~Salvo T.,  Papitto A.,  Marino A.,  Iaria R.,   Burderi L.,  2024,
  Low-Magnetic-Field Neutron Stars in X-ray Binaries.
Springer Nature Singapore, Singapore, pp 4031--4103,
  \mn@doi{10.1007/978-981-19-6960-7_103}, \url
  {https://doi.org/10.1007/978-981-19-6960-7_103}

\bibitem[\protect\citeauthoryear{Dubus, Lasota, Hameury  \& Charles}{Dubus
  et~al.}{1999}]{Dubus1999}
Dubus G.,  Lasota J.~P.,  Hameury J.~M.,   Charles P.,  1999, \mn@doi [Monthly
  Notices of the Royal Astronomical Society]
  {10.1046/J.1365-8711.1999.02212.X/2/303-1-139-FIG010.JPEG}, 303, 139

\bibitem[\protect\citeauthoryear{{Eckert}, {Walter}, {Kretschmar}, {Mas-Hesse},
  {Palumbo}, {Roques}, {Ubertini}  \& {Winkler}}{{Eckert}
  et~al.}{2004}]{Eckert2004}
{Eckert} D.,  {Walter} R.,  {Kretschmar} P.,  {Mas-Hesse} M.,  {Palumbo}
  G.~G.~C.,  {Roques} J.-P.,  {Ubertini} P.,   {Winkler} C.,  2004, The
  Astronomer's Telegram, \href
  {https://ui.adsabs.harvard.edu/abs/2004ATel..352....1E} {352, 1}

\bibitem[\protect\citeauthoryear{Ertan}{Ertan}{2017}]{Ertan2017}
Ertan {\"U}.,  2017, \mn@doi [Monthly Notices of the Royal Astronomical
  Society] {10.1093/MNRAS/STW3131}, 466, 175

\bibitem[\protect\citeauthoryear{Ertan}{Ertan}{2018}]{Ertan2018}
Ertan {\"U}.,  2018, \mn@doi [Monthly Notices of the Royal Astronomical
  Society: Letters] {10.1093/MNRASL/SLY089}, 479, L12

\bibitem[\protect\citeauthoryear{Ertan}{Ertan}{2021}]{Ertan2020}
Ertan {\"U}.,  2021, \mn@doi [Monthly Notices of the Royal Astronomical
  Society] {10.1093/MNRAS/STAA3378}, 500, 2928

\bibitem[\protect\citeauthoryear{Ertan \& Alpar}{Ertan \&
  Alpar}{2021}]{ertan2021}
Ertan {\"U}.,  Alpar M.~A.,  2021, \mn@doi [Monthly Notices of the Royal
  Astronomical Society: Letters] {10.1093/mnrasl/slab060}, 505, L112

\bibitem[\protect\citeauthoryear{Falanga et~al.,}{Falanga
  et~al.}{2005}]{Falanga2005}
Falanga M.,  et~al., 2005, \mn@doi [Astronomy and Astrophysics]
  {10.1051/0004-6361:20053472}, 444, 15

\bibitem[\protect\citeauthoryear{Frank, King  \& Raine}{Frank
  et~al.}{2002}]{Frank2002}
Frank J.,  King A.,   Raine D.~J.,  2002, Cambridge University Press, 39, 398

\bibitem[\protect\citeauthoryear{Galloway, Markwardt, Morgan, Chakrabarty  \&
  Strohmayer}{Galloway et~al.}{2005}]{Galloway2005}
Galloway D.~K.,  Markwardt C.~B.,  Morgan E.~H.,  Chakrabarty D.,   Strohmayer
  T.~E.,  2005, \mn@doi [The Astrophysical Journal] {10.1086/429563}, 622, L45

\bibitem[\protect\citeauthoryear{Gençali et~al.,}{Gençali
  et~al.}{2022}]{Genali2022}
Gençali A.~A.,  et~al., 2022, \mn@doi [Astronomy and Astrophysics]
  {10.1051/0004-6361/202141772}, 658, A13

\bibitem[\protect\citeauthoryear{Gierliński \& Poutanen}{Gierliński \&
  Poutanen}{2005}]{Gierliski2005}
Gierliński M.,  Poutanen J.,  2005, \mn@doi [Monthly Notices of the Royal
  Astronomical Society] {10.1111/j.1365-2966.2005.09004.x}, 359, 1261

\bibitem[\protect\citeauthoryear{{Hartman} et~al.,}{{Hartman}
  et~al.}{2008}]{Hartman2008}
{Hartman} J.~M.,  et~al., 2008, \mn@doi [\apj] {10.1086/527461}, \href
  {https://ui.adsabs.harvard.edu/abs/2008ApJ...675.1468H} {675, 1468}

\bibitem[\protect\citeauthoryear{Haskell \& Patruno}{Haskell \&
  Patruno}{2011}]{Haskell_2011}
Haskell B.,  Patruno A.,  2011, \mn@doi [The Astrophysical Journal]
  {10.1088/2041-8205/738/1/l14}, 738, L14

\bibitem[\protect\citeauthoryear{{Illiano} et~al.,}{{Illiano}
  et~al.}{2025}]{Illiano2025}
{Illiano} G.,  et~al., 2025, \mn@doi [arXiv e-prints]
  {10.48550/arXiv.2507.13248}, \href
  {https://ui.adsabs.harvard.edu/abs/2025arXiv250713248I} {p. arXiv:2507.13248}

\bibitem[\protect\citeauthoryear{Jonker, Campana, Steeghs, Torres, Galloway,
  Markwardt, Chakrabarty  \& Swank}{Jonker et~al.}{2005}]{Jonker2005}
Jonker P.~G.,  Campana S.,  Steeghs D.,  Torres M.~A.,  Galloway D.~K.,
  Markwardt C.~B.,  Chakrabarty D.,   Swank J.,  2005, \mn@doi [Monthly Notices
  of the Royal Astronomical Society] {10.1111/j.1365-2966.2005.09171.x}, 361,
  511

\bibitem[\protect\citeauthoryear{{Krimm} et~al.,}{{Krimm}
  et~al.}{2007}]{Krimm2007}
{Krimm} H.~A.,  et~al., 2007, \mn@doi [\apjl] {10.1086/522959}, \href
  {https://ui.adsabs.harvard.edu/abs/2007ApJ...668L.147K} {668, L147}

\bibitem[\protect\citeauthoryear{Lasota}{Lasota}{2001}]{Lasota2001}
Lasota J.~P.,  2001, \mn@doi [New Astronomy Reviews]
  {10.1016/S1387-6473(01)00112-9}, 45, 449

\bibitem[\protect\citeauthoryear{{Li} et~al.,}{{Li} et~al.}{2021}]{Li2021}
{Li} Z.~S.,  et~al., 2021, \mn@doi [\aap] {10.1051/0004-6361/202140360}, \href
  {https://ui.adsabs.harvard.edu/abs/2021A&A...649A..76L} {649, A76}

\bibitem[\protect\citeauthoryear{Lovelace, Romanova  \&
  Bisnovatyi-Kogan}{Lovelace et~al.}{1995}]{Lovelace1995}
Lovelace R. V.~E.,  Romanova M.~M.,   Bisnovatyi-Kogan G.~S.,  1995, \mn@doi
  [Monthly Notices of the Royal Astronomical Society]
  {10.1093/mnras/275.2.244}, 275, 244

\bibitem[\protect\citeauthoryear{Lovelace, Romanova  \&
  Bisnovatyi-Kogan}{Lovelace et~al.}{1999}]{Lovelace_1999}
Lovelace R. V.~E.,  Romanova M.~M.,   Bisnovatyi-Kogan G.~S.,  1999, \mn@doi
  [The Astrophysical Journal] {10.1086/306945}, 514, 368

\bibitem[\protect\citeauthoryear{{Markwardt}, {Swank}, {Strohmayer}, {in 't
  Zand}  \& {Marshall}}{{Markwardt} et~al.}{2002}]{Markwardt2002}
{Markwardt} C.~B.,  {Swank} J.~H.,  {Strohmayer} T.~E.,  {in 't Zand} J.~J.~M.,
    {Marshall} F.~E.,  2002, \mn@doi [\apjl] {10.1086/342612}, \href
  {https://ui.adsabs.harvard.edu/abs/2002ApJ...575L..21M} {575, L21}

\bibitem[\protect\citeauthoryear{{Markwardt}, {Swank}  \&
  {Strohmayer}}{{Markwardt} et~al.}{2004}]{Markwardt2004}
{Markwardt} C.~B.,  {Swank} J.~H.,   {Strohmayer} T.~E.,  2004, The
  Astronomer's Telegram, \href
  {https://ui.adsabs.harvard.edu/abs/2004ATel..353....1M} {353, 1}

\bibitem[\protect\citeauthoryear{{Markwardt}, {Altamirano}, {Strohmayer}  \&
  {Swank}}{{Markwardt} et~al.}{2009}]{Markwardt2009}
{Markwardt} C.~B.,  {Altamirano} D.,  {Strohmayer} T.~E.,   {Swank} J.~H.,
  2009, The Astronomer's Telegram, \href
  {https://ui.adsabs.harvard.edu/abs/2009ATel.2237....1M} {2237, 1}

\bibitem[\protect\citeauthoryear{{Meyer} \& {Meyer-Hofmeister}}{{Meyer} \&
  {Meyer-Hofmeister}}{1984}]{meyer1984}
{Meyer} F.,  {Meyer-Hofmeister} E.,  1984, \aap, \href
  {https://ui.adsabs.harvard.edu/abs/1984A&A...132..143M} {132, 143}

\bibitem[\protect\citeauthoryear{Mukherjee, Bult, Derklis  \&
  Bhattacharya}{Mukherjee et~al.}{2015}]{Mukherjee2015}
Mukherjee D.,  Bult P.,  Derklis M.~V.,   Bhattacharya D.,  2015, \mn@doi
  [Monthly Notices of the Royal Astronomical Society] {10.1093/mnras/stv1542},
  452, 3994

\bibitem[\protect\citeauthoryear{{Ng} et~al.,}{{Ng} et~al.}{2020}]{Ng2020}
{Ng} M.,  et~al., 2020, The Astronomer's Telegram, \href
  {https://ui.adsabs.harvard.edu/abs/2020ATel14124....1N} {14124, 1}

\bibitem[\protect\citeauthoryear{Ng et~al.,}{Ng et~al.}{2021}]{Ng2021}
Ng M.,  et~al., 2021, \mn@doi [The Astrophysical Journal Letters]
  {10.3847/2041-8213/abe1b4}, 908, L15

\bibitem[\protect\citeauthoryear{Niang, Ertan, Gençali, Toyran, Ulubay,
  Devlen, Alpar  \& Gügercinoǧlu}{Niang et~al.}{2024}]{Niang2024}
Niang N.,  Ertan {\"U}.,  Gençali A.~A.,  Toyran O.,  Ulubay A.,  Devlen E.,
  Alpar M.~A.,   Gügercinoǧlu E.,  2024, \mn@doi [Monthly Notices of the
  Royal Astronomical Society] {10.1093/mnras/stae1595}, 532, 2133

\bibitem[\protect\citeauthoryear{{Niang}, {Ertan}, {Gen{\c{c}}ali},
  {Ertu{\u{g}}rul}, {Ulubay}, {Devlen}  \& {Alpar}}{{Niang}
  et~al.}{2026}]{niang2026}
{Niang} N.,  {Ertan} {\"U}.,  {Gen{\c{c}}ali} A.~A.,  {Ertu{\u{g}}rul} F.,
  {Ulubay} A.,  {Devlen} E.,   {Alpar} M.~A.,  2026, \mn@doi [\mnras]
  {10.1093/mnras/stag745}, \href
  {https://ui.adsabs.harvard.edu/abs/2026MNRAS.548ag745N} {548, stag745}

\bibitem[\protect\citeauthoryear{Papitto \& Martino}{Papitto \&
  Martino}{2022}]{Papitto2022}
Papitto A.,  Martino D.~d.,  2022, in Bhattacharyya S.,  Papitto A.,
  Bhattacharya D.,  eds, , Millisecond Pulsars.
Springer International Publishing, Cham, pp 157--200,
  \mn@doi{10.1007/978-3-030-85198-9_6}, \url
  {https://doi.org/10.1007/978-3-030-85198-9_6}

\bibitem[\protect\citeauthoryear{{Papitto} et~al.,}{{Papitto}
  et~al.}{2007}]{Papitto2007}
{Papitto} A.,  et~al., 2007, The Astronomer's Telegram, \href
  {https://ui.adsabs.harvard.edu/abs/2007ATel.1133....1P} {1133, 1}

\bibitem[\protect\citeauthoryear{Papitto, Menna, Burderi, Salvo  \&
  Riggio}{Papitto et~al.}{2008}]{Papitto2008}
Papitto A.,  Menna M.~T.,  Burderi L.,  Salvo T.~D.,   Riggio A.,  2008,
  \mn@doi [Monthly Notices of the Royal Astronomical Society]
  {10.1111/j.1365-2966.2007.12551.x}, 383, 411

\bibitem[\protect\citeauthoryear{Papitto, Riggio, Burderi, Salvo, D'Aí  \&
  Iaria}{Papitto et~al.}{2011}]{Papitto2011}
Papitto A.,  Riggio A.,  Burderi L.,  Salvo T.~D.,  D'Aí A.,   Iaria R.,
  2011, \mn@doi [Astronomy and Astrophysics] {10.1051/0004-6361/201014837}, 528

\bibitem[\protect\citeauthoryear{{Papitto} et~al.,}{{Papitto}
  et~al.}{2013}]{papitto2013}
{Papitto} A.,  et~al., 2013, \mn@doi [\nat] {10.1038/nature12470}, \href
  {https://ui.adsabs.harvard.edu/abs/2013Natur.501..517P} {501, 517}

\bibitem[\protect\citeauthoryear{Patruno}{Patruno}{2010}]{Patruno2010}
Patruno A.,  2010, \mn@doi [The Astrophysical Journal]
  {10.1088/0004-637X/722/1/909}, 722, 909

\bibitem[\protect\citeauthoryear{Patruno \& Watts}{Patruno \&
  Watts}{2021}]{Patruno2021}
Patruno A.,  Watts A.~L.,  2021, in Belloni T.~M.,  M{\'e}ndez M.,   Zhang C.,
  eds, , Timing Neutron Stars: Pulsations, Oscillations and Explosions.
Springer Berlin Heidelberg, Berlin, Heidelberg, pp 143--208,
  \mn@doi{10.1007/978-3-662-62110-3_4}, \url
  {https://doi.org/10.1007/978-3-662-62110-3_4}

\bibitem[\protect\citeauthoryear{Patruno, Altamirano  \& Messenger}{Patruno
  et~al.}{2010a}]{patruno2010b}
Patruno A.,  Altamirano D.,   Messenger C.,  2010a, \mn@doi [Monthly Notices of
  the Royal Astronomical Society] {10.1111/j.1365-2966.2010.16202.x}, 403, 1426

\bibitem[\protect\citeauthoryear{Patruno, Altamirano  \& Messenger}{Patruno
  et~al.}{2010b}]{Patruno10}
Patruno A.,  Altamirano D.,   Messenger C.,  2010b, \mn@doi [Monthly Notices of
  the Royal Astronomical Society] {10.1111/j.1365-2966.2010.16202.x}, 403, 1426

\bibitem[\protect\citeauthoryear{{Radhakrishnan} \&
  {Srinivasan}}{{Radhakrishnan} \& {Srinivasan}}{1982}]{Radhakrishnan1982}
{Radhakrishnan} V.,  {Srinivasan} G.,  1982, Current Science, \href
  {https://ui.adsabs.harvard.edu/abs/1982CSci...51.1096R} {51, 1096}

\bibitem[\protect\citeauthoryear{{Riggio}, {Papitto}, {Burderi}, {di Salvo},
  {Bachetti}, {Iaria}, {D'A{\`\i}}  \& {Menna}}{{Riggio}
  et~al.}{2011a}]{Riggio2011b}
{Riggio} A.,  {Papitto} A.,  {Burderi} L.,  {di Salvo} T.,  {Bachetti} M.,
  {Iaria} R.,  {D'A{\`\i}} A.,   {Menna} M.~T.,  2011a, \mn@doi [\aap]
  {10.1051/0004-6361/201014322}, \href
  {https://ui.adsabs.harvard.edu/abs/2011A&A...526A..95R} {526, A95}

\bibitem[\protect\citeauthoryear{{Riggio}, {Burderi}, {di Salvo}, {Papitto},
  {D'A{\`\i}}, {Iaria}  \& {Menna}}{{Riggio} et~al.}{2011b}]{Riggio2011a}
{Riggio} A.,  {Burderi} L.,  {di Salvo} T.,  {Papitto} A.,  {D'A{\`\i}} A.,
  {Iaria} R.,   {Menna} M.~T.,  2011b, \mn@doi [\aap]
  {10.1051/0004-6361/201014883}, \href
  {https://ui.adsabs.harvard.edu/abs/2011A&A...531A.140R} {531, A140}

\bibitem[\protect\citeauthoryear{Sanna et~al.,}{Sanna et~al.}{2017}]{Sanna2017}
Sanna A.,  et~al., 2017, \mn@doi [Monthly Notices of the Royal Astronomical
  Society] {10.1093/mnras/stw3332}, 466, 2910

\bibitem[\protect\citeauthoryear{Sanna et~al.,}{Sanna et~al.}{2018}]{Sanna2018}
Sanna A.,  et~al., 2018, \mn@doi [Monthly Notices of the Royal Astronomical
  Society] {10.1093/MNRAS/STY2316}, 481, 1658

\bibitem[\protect\citeauthoryear{{Sanna} et~al.,}{{Sanna}
  et~al.}{2025}]{sanna2025}
{Sanna} A.,  et~al., 2025, \mn@doi [\aap] {10.1051/0004-6361/202555734}, \href
  {https://ui.adsabs.harvard.edu/abs/2025A&A...703A.171S} {703, A171}

\bibitem[\protect\citeauthoryear{{Sguera}}{{Sguera}}{2025}]{Sguera2025}
{Sguera} V.,  2025, The Astronomer's Telegram, \href
  {https://ui.adsabs.harvard.edu/abs/2025ATel17029....1S} {17029, 1}

\bibitem[\protect\citeauthoryear{{Shakura} \& {Sunyaev}}{{Shakura} \&
  {Sunyaev}}{1973}]{Shakura1973}
{Shakura} N.~I.,  {Sunyaev} R.~A.,  1973, \aap, \href
  {http://adsabs.harvard.edu/abs/1973A%26A....24..337S} {24, 337}

\bibitem[\protect\citeauthoryear{{Takagi}, {Mihara}, {Sugizaki}, {Makishima}
  \& {Morii}}{{Takagi} et~al.}{2016}]{Takagi2016}
{Takagi} T.,  {Mihara} T.,  {Sugizaki} M.,  {Makishima} K.,   {Morii} M.,
  2016, \mn@doi [\pasj] {10.1093/pasj/psw010}, \href
  {https://ui.adsabs.harvard.edu/abs/2016PASJ...68S..13T} {68, S13}

\bibitem[\protect\citeauthoryear{Ustyugova, Koldoba, Romanova  \&
  Lovelace}{Ustyugova et~al.}{2006}]{Ustyugova_2006}
Ustyugova G.~V.,  Koldoba A.~V.,  Romanova M.~M.,   Lovelace R. V.~E.,  2006,
  \mn@doi [The Astrophysical Journal] {10.1086/503379}, 646, 304

\bibitem[\protect\citeauthoryear{{Wijnands}, {Homan}, {Heinke}, {Miller}  \&
  {Lewin}}{{Wijnands} et~al.}{2005}]{Wijnands2005}
{Wijnands} R.,  {Homan} J.,  {Heinke} C.~O.,  {Miller} J.~M.,   {Lewin} W.
  H.~G.,  2005, \mn@doi [\apj] {10.1086/426379}, \href
  {https://ui.adsabs.harvard.edu/abs/2005ApJ...619..492W} {619, 492}

\bibitem[\protect\citeauthoryear{{van Paradijs} \& {McClintock}}{{van Paradijs}
  \& {McClintock}}{1994}]{van1994}
{van Paradijs} J.,  {McClintock} J.~E.,  1994, \aap, \href
  {https://ui.adsabs.harvard.edu/abs/1994A&A...290..133V} {290, 133}

\makeatother
\end{thebibliography}






\bsp	
\label{lastpage}
\end{document}